\documentclass[a4paper]{jpconf}

\usepackage{graphicx}
\usepackage{amsmath,amssymb}
\usepackage{subfigure}
\usepackage{color}
\usepackage{multirow}
\usepackage{cite}

\begin{document}

\title{Frustrated Heisenberg antiferromagnet on the honeycomb lattice with spin quantum number $s \geq 1$}


\author{P H Y Li$^1$, \underline{R F Bishop}$^1$ 
and C E Campbell$^2$}

\address{$^1$ School of Physics and Astronomy, Schuster Building, The University of Manchester, Manchester, M13 9PL, UK}
\address{$^2$ School of Physics and Astronomy, University of Minnesota, 116 Church Street SE, Minneapolis, Minnesota 55455, USA}

\ead{raymond.bishop@manchester.ac.uk; peggyhyli@gmail.com}

\begin{abstract}
  The ground-state (GS) phase diagram of the frustrated spin-$s$
  $J_{1}$--$J_{2}$--$J_{3}$ Heisenberg antiferromagnet on the
  honeycomb lattice is studied using the coupled cluster method
  implemented to high orders of approximation, for spin quantum
  numbers $s=1,\,\frac{3}{2},\,2\,,\frac{5}{2}$.  The model has
  antiferromagnetic (AFM) nearest-neighbour, next-nearest-neighbour and
  next-next-nearest-neighbour exchange couplings (with strength
  $J_{1}>0$, $J_{2}>0$ and $J_{3}>0$, respectively). We specifically
  study the case $J_{3}=J_{2}=\kappa J_{1}$, in the range $0 \leq
  \kappa \leq 1$ of the frustration parameter, which includes the
  point of maximum classical ($s \rightarrow \infty$) frustration,
  viz., the classical critical point at $\kappa_{{\rm
      cl}}=\frac{1}{2}$, which separates the N\'{e}el phase for
  $\kappa < \kappa_{{\rm cl}}$ and the collinear striped AFM phase for
  $\kappa > \kappa_{{\rm cl}}$.  Results are presented for the GS
  energy, magnetic order parameter and plaquette valence-bond crystal
  (PVBC) susceptibility.  For all spins $s \geq \frac{3}{2}$ we find a
  quantum phase diagram very similar to the classical one, with a
  direct first-order transition between the two collinear AFM states
  at a value $\kappa_{c}(s)$ which is slightly greater than
  $\kappa_{{\rm cl}}$ [e.g., $\kappa_{c}(\frac{3}{2}) \approx
  0.53(1)$] and which approaches it monotonically as $s \rightarrow
  \infty$.  By contrast, for the case $s=1$ the transition is split
  into two such that the stable GS phases are one with N\'{e}el AFM
  order for $\kappa < \kappa_{c_{1}} = 0.485(5)$ and one with striped
  AFM order for $\kappa > \kappa_{c_{2}} = 0.528(5)$, just as in the
  case $s=\frac{1}{2}$ (for which $\kappa_{c_{1}} \approx 0.47$ and
  $\kappa_{c_{2}} \approx 0.60$).  For both the $s=\frac{1}{2}$ and $s=1$ models
  the transition at $\kappa_{c_{2}}$ appears to be of first-order
  type, while that at $\kappa_{c_{1}}$ appears to be continuous.
  However, whereas in the $s=\frac{1}{2}$ case the intermediate phase
  appears to have PVBC order over the entire range $\kappa_{c_{1}} <
  \kappa < \kappa_{c_{2}}$, in the $s=1$ case PVBC ordering either
  exists only over a very small part of the region or, more likely, is
  absent everywhere.
\end{abstract}

\section{Introduction}
\label{intro}
Quantum spin-lattice models, in which the sites of a give regular
periodic lattice are all occupied by magnetic ions with spin quantum
number $s$, offer a rich arena for the study of exotic ground-state
(GS) phases that are not present in their classical ($s \rightarrow
\infty$) counterparts.  Whereas interactions between the classical
spins give rise to magnetic ground states in which the spins are ordered such
that each individual spin is oriented in a specific direction, quantum
fluctuations can act either to diminish the corresponding magnetic
order parameter (viz., the average local onsite magnetization) or to
destroy it altogether.  In the former case, where the long-range order
(LRO) is only partially reduced, such quasiclassical magnetically
ordered states spontaneously break both the SU(2) spin-rotation and
time-reversal symmetries. 

By contrast, such intrinsically quantum-mechanical states as the
various valence-bond crystalline (VBC) phases, in which specific
combinations of the lattice spins combine into spin singlets, have
zero magnetic order and break neither of the SU(2) spin-rotation and
time-reversal symmetries, although they still break some lattice
symmetries.  Yet other states exist in which, for example,
time-reversal symmetry is preserved, so that magnetic order is
definitely absent, but for which the SU(2) spin-rotation symmetry is
still broken.  These are the so-called multipolar or spin-nematic
phases.  Finally, of course, one also has the possibility of quantum
spin-liquid (QSL) phases that preserve all of the symmetries,
including the lattice symmetries.

For a given regular lattice in $d$ spatial dimensions, one is usually
interested in the interplay between frustration (which may be tuned,
for example, by varying the relative strengths of competing
interactions in the model Hamiltonian that separately tend to promote
different forms of magnetic LRO) and quantum fluctuations.  In broad
terms quantum fluctuations are larger for lower values of both the
spatial dimensionality $d$ and the spin quantum number $s$.  They are
typically also larger, for given spatial dimensionality $d$, for
lattices with smaller values of the coordination number $z$.

For the case of $d=1$ isotropic Heisenberg chain systems the
Mermin-Wagner theorem \cite{Mermin:1966} excludes the possibility of GS
magnetic order even at zero temperature ($T=0$), since it is
impossible to break a continuous symmetry for any such system.  The
Mermin-Wagner theorem similarly implies the absence of magnetic LRO in
any $d=2$ isotropic Heisenberg system at all nonzero temperatures ($T
> 0$).  The behaviour and GS quantum phase structure of two-dimensional (2D)
spin-lattice models at $T=0$ has thus come to occupy a special role in
the study of quantum phase transitions.

Spin-lattice systems are said to be frustrated when constraints are
present that preclude the formation of a GS phase which satisfies all
of the (generally, pairwise) interactions among the spins.
Frustration is strongly associated with macroscopic degeneracy of the
GS phase, with the consequent existence of strong quantum fluctuations
among the states in the degenerate manifold.  Either quantum or
thermal fluctuations can then, in such a situation, suppress magnetic
LRO, and the possibility of such exotic non-classical states as those
discussed above forming the stable GS phase under certain conditions
is heightened.  The low coordination number, $z=3$, of the honeycomb
lattice further enhances the quantum fluctuations, and thereby makes
it a special 2D spin-lattice model candidate for the study of its
$T=0$ GS quantum phase diagram when dynamical frustration is introduced via
competing interactions.

In the present paper we study the so-called $J_{1}$--$J_{2}$--$J_{3}$
model on the honeycomb lattice with antiferromagnetic (AFM) Heisenberg
exchange interactions between pairs of nearest-neighbour (NN) spins
(of strength $J_{1}$), next-nearest-neighbour (NNN) spins (of strength
$J_{2}$), and next-next-nearest-neighbour (NNNN) spins (of strength
$J_{3}$).  Even in the classical ($s \rightarrow \infty$) limit the
model has a rich phase diagram
\cite{Rastelli:1979_honey,Fouet:2001_honey}, as we discuss further in
Sec.\ \ref{model_sec} below.  In the case where all three bonds are
AFM in nature (i.e., $J_{i} > 0$; $i = 1,2,3$), the classical system
exhibits two collinear AFM phases, namely the so-called N\'{e}el and
striped phases, as well as a spiral phase.  The three phases meet in a
triple point at $J_{3} = J_{2} = J_{1}/2$ (see, e.g., Refs.\
\cite{Fouet:2001_honey,Cabra:2011_honey}).  This is the point of
maximum classical frustration, where the classical GS phase has
macroscopic degeneracy.  For the present study we consider the model
along the line $J_{3}=J_{2} \equiv \kappa J_{1}$ for the case
$J_{1}>0$, as a function of the frustration parameter $\kappa$ in the
range $0 \leq \kappa \leq 1$.  The classical ($s \rightarrow \infty$)
version of the model thus has a single quantum phase transition in its
$T=0$ phase diagram at $\kappa_{{\rm cl}} = \frac{1}{2}$.  For $\kappa
< \kappa_{{\rm cl}}$ the stable GS phase is the N\'{e}el AFM phase,
whereas for $\kappa > \kappa_{{\rm cl}}$ the stable GS phase is the
striped AFM phase.  In fact, at $T=0$, there actually exists an
infinite family of non-coplanar states, all of which are degenerate in
energy with the striped state.  However, it is asserted \cite{Fouet:2001_honey}
that both thermal and quantum fluctuations break this degeneracy in
favour of the collinear striped state, at least in the large-$s$ limit \cite{Rastelli:1979_honey}.

Whereas the spin-$\frac{1}{2}$ $J_{1}$--$J_{2}$--$J_{3}$ model on the
honeycomb lattice, or particular cases of it (e.g., when $J_{3}=J_{2}$
or $J_{3}=0$), have been investigated by many authors with a variety
of theoretical tools (see, e.g., Refs.\
\cite{Rastelli:1979_honey,Fouet:2001_honey,Mattsson:1994_honey,Mulder:2010_honey,Cabra:2011_honey,Ganesh:2011_honey,Clark:2011_honey,DJJF:2011_honeycomb,Reuther:2011_honey,Albuquerque:2011_honey,Mosadeq:2011_honey,Oitmaa:2011_honey,Mezzacapo:2012_honey,PHYLi:2012_honeycomb_J1neg,Bishop:2012_honeyJ1-J2,Bishop:2012_honey_phase,Li:2012_honey_full,RFB:2013_hcomb_SDVBC,Ganesh:2013_honey_J1J2mod-XXX,Zhu:2013_honey_XY,Zhang:2013_honey,Gong:2013_J1J2mod-XXX,Yu:2014_honey_J1J2mod}),
there are far fewer studies of the model in the case $s>\frac{1}{2}$.
A particular exception is a very recent study
\cite{Gong:2015_honey_J1J2mod_s1} of the $s=1$ $J_{1}$--$J_{2}$ model
(i.e., when $J_{3}=0$) on the honeycomb lattice, using the
density-matrix renormalization group (DMRG) method.  Our specific aim
here is to extend earlier work using the coupled cluster method (CCM)
applied to the spin-$\frac{1}{2}$ version of the
$J_{1}$--$J_{2}$--$J_{3}$ model on the honeycomb lattice along the
line $J_{3}=J_{2}$
\cite{DJJF:2011_honeycomb,PHYLi:2012_honeycomb_J1neg,Bishop:2012_honey_phase},
to cases $s>\frac{1}{2}$.  In particular, we now compare results for
the case $s=\frac{1}{2}$ with those for
$s=1,\,\frac{3}{2},\,2,\,\frac{5}{2}$.

We note that by now there exist many experimental realizations of
frustrated honeycomb-lattice systems with AFM interactions.  These
include such magnetic compounds as Na$_{3}$Cu$_{2}$SbO$_{6}$
\cite{Miura:2006_honey}, InCu$_{2/3}$V$_{1/3}$O$_{3}$
\cite{Kataev:2005_honey}, $\beta$-Cu$_{2}$V$_{2}$O$_{7}$
\cite{Tsirlin:2010_honey}, and Cu$_{5}$SbO$_{6}$
\cite{Climent:2012_honey}, in each of which the $s=\frac{1}{2}$
Cu$^{2+}$ ions are situated on the sites of weakly coupled
honeycomb-lattice layers.  The iridates A$_{2}$IrO$_{3}$ (A $=$ Na, Li)
\cite{Singh:2010_honey,Liu:2011_honey,Singh:2012_honey,Choi:2012_honey}
are also believed to be magnetically ordered Mott insulators in which
the Ir$^{4+}$ ions form effective $s=\frac{1}{2}$ moments arrayed on
weakly-coupled honeycomb-lattice layers.  Other similar honeycomb
materials include, for example, the families of compounds
BaM$_{2}$(XO$_{4}$)$_{2}$ (M $=$ Co, Ni; X $=$ P, As)
\cite{Regnault:1990_honey} and Cu$_{3}$M$_{2}$SbO$_{6}$ (M $=$ Co, Ni)
\cite{Roudebush:2013_honey}.  In both of these families the magnetic
ions M$^{2+}$ are again disposed in weakly-coupled layers where they
occupy the sites of a honeycomb lattice.  In both families the
Ni$^{2+}$ ions appear to take the high-spin value $s=1$, whereas the
Co$^{2+}$ ions appear to take the low-spin value $s=\frac{1}{2}$ in
the former family BaCo$_{2}$(XO$_{4}$)$_{2}$ and the high-spin value
$s=\frac{3}{2}$ in the latter compound Cu$_{3}$Co$_{2}$SbO$_{6}$.  As a
last example of an $s=\frac{3}{2}$ honeycomb-lattice AFM material, we
also mention the layered compound Bi$_{3}$Mn$_{4}$O$_{12}$(NO$_{3}$)
\cite{Smirnova:2009:honey_spin_3half,Okubo:2010:honey_spin_3half} in
which the spin-$\frac{3}{2}$ Mn$^{4+}$ ions sit on the sites of
the honeycomb layers.

The remainder of the paper is organized as follows.  In Sec.\
\ref{model_sec} we discuss further the model itself, before we give a
brief description in Sec.\ \ref{ccm_section} of the CCM formalism that
we apply to it.  The results are then presented in Sec.\
\ref{results_sec}, and we conclude in Sec.\ \ref{conclusions_sec} with a
discussion and summary.

\section{The model}
\label{model_sec}
\begin{figure}[t]
\begin{center}
\mbox{
\subfigure[]{\includegraphics[width=4.0cm]{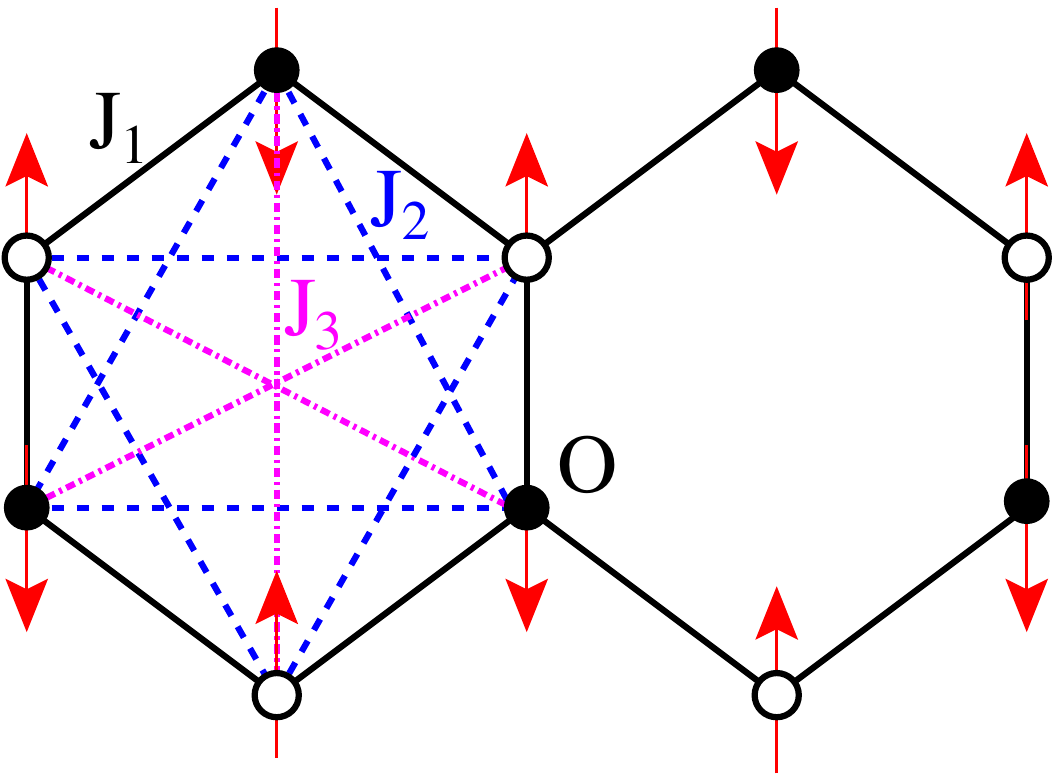}}
\quad \subfigure[]{\includegraphics[width=4.0cm]{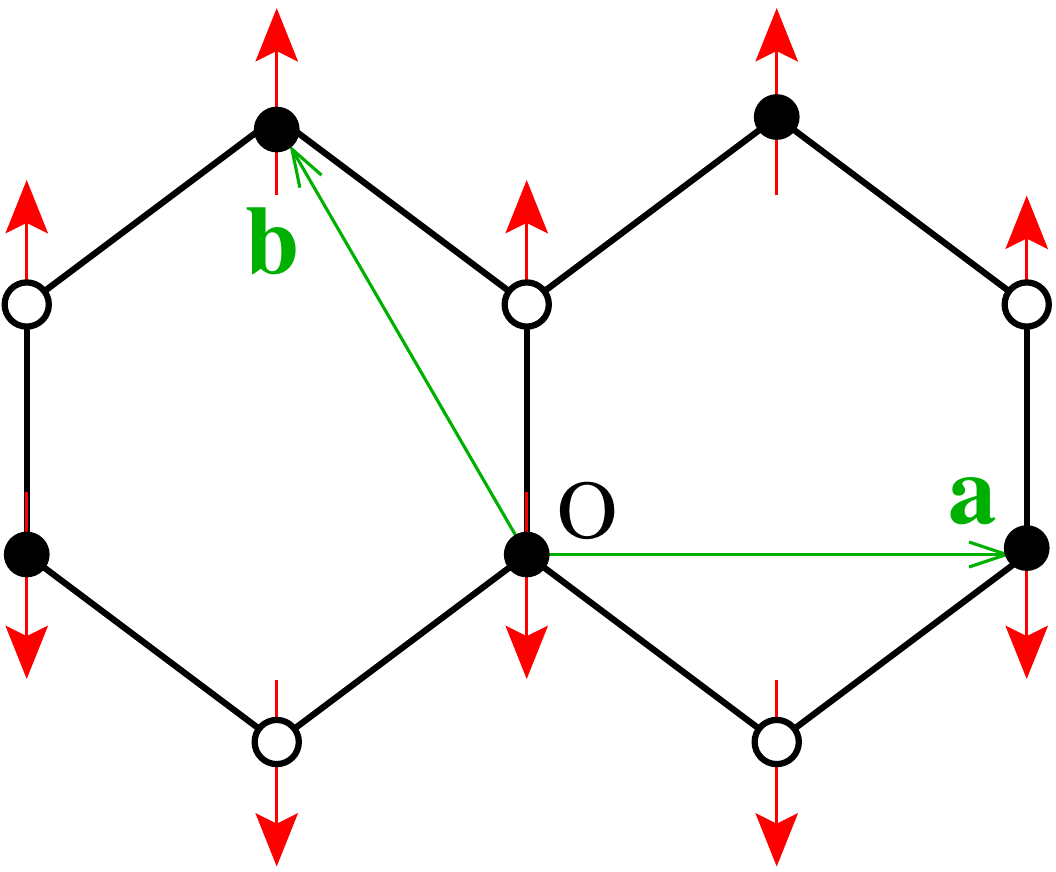}}
\quad \subfigure{\includegraphics[width=2.5cm]{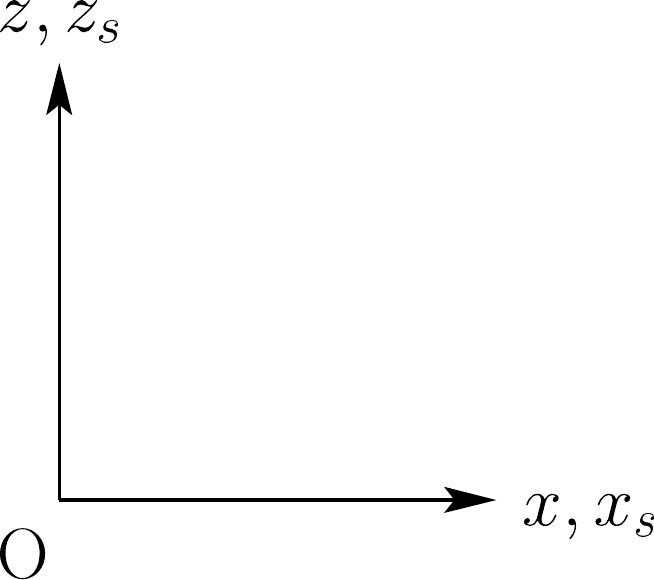}}
  }
  \caption{The $J_{1}$--$J_{2}$--$J_{3}$ model on the honeycomb
    lattice, showing (a) the bonds ($J_{1} = $ -----, $J_{2} = $ - - -,
    $J_{3} = $ - $\cdot$ - ) and the N\'{e}el state, and (b) the
    triangular Bravais lattice vectors $\mathbf{a}$ and $\mathbf{b}$
    and one of three equivalent striped states.  Sites on the two
    triangular sublattices ${\cal A}$ and ${\cal B}$ are shown by
    filled and empty circles respectively, and the spins are
    represented by the (red) arrows on the lattice sites.}
\label{model_pattern}
\end{center}
\end{figure}

The Hamiltonian of the $J_{1}$--$J_{2}$--$J_{3}$ model on the honeycomb lattice is given by
\begin{equation}
H = J_{1}\sum_{\langle i,j \rangle} \mathbf{s}_{i}\cdot\mathbf{s}_{j} + J_{2}\sum_{\langle\langle i,k \rangle\rangle} \mathbf{s}_{i}\cdot\mathbf{s}_{k} + J_{3}\sum_{\langle\langle\langle i,l \rangle\rangle\rangle} \mathbf{s}_{i}\cdot\mathbf{s}_{l}\,,
\label{Hamiltonian}
\end{equation}
where the sums over $\langle i,j \rangle$, $\langle \langle i,k \rangle
\rangle$ and $\langle \langle \langle i,l \rangle \rangle \rangle$ run
over all NN, NNN and NNNN bonds, respectively, on the lattice,
counting each pair of spins once and once only in each of the three
sums.  Each site $i$ of the honeycomb lattice carries a spin-$s$
particle described by the SU(2) spin operator ${\bf s}_{i} \equiv
(s^{x}_{i}, s^{y}_{i}, s^{z}_{i})$, with ${\bf s}^{2}_{i} = s(s+1)$,
and, for the cases considered here,
$s=1,\,\frac{3}{2},\,2,\frac{5}{2}$.  The lattice and the Heisenberg
exchange bonds are illustrated in Fig.~\ref{model_pattern}(a).
For the present study we are interested in the case where each of the
three types of bonds is AFM in nature (i.e., $J_{m} > 0$; $m=1,2,3$).
Without loss of generality we may put $J_{1} \equiv 1$ to set the
overall energy scale, and we will specifically consider the case where
$J_{3}=J_{2} \equiv \kappa J_{1}$, in the interesting window $0 \leq \kappa \leq 1$ of the frustration parameter $\kappa$.

The honeycomb lattice is bipartite, but comprises two triangular
Bravais sublattices ${\cal A}$ and ${\cal B}$.  The basis vectors
$\mathbf{a}=\sqrt{3}d\hat{x}$ and
$\mathbf{b}=(-\sqrt{3}\hat{x}+3\hat{z})d/2$ are illustrated in Fig.\
\ref{model_pattern}(b), where the lattice is defined to lie in the
$xz$ plane as shown, and where $d$ is the lattice spacing (i.e., the
distance between NN sites).  The unit cell $i$ at position vector
$\mathbf{R}_{i} = m_{i}\mathbf{a}+n_{i}\mathbf{b}$, where $m_{i},\,n_{i} \in
\mathbb{Z}$, now comprises the two sites at $\mathbf{R}_{i} \in {\cal
  A}$ and $(\mathbf{R}_{i}+d\hat{z}) \in {\cal B}$.  The reciprocal
lattice vectors corresponding to the real-space vectors $\mathbf{a}$
and $\mathbf{b}$ are thus
$\boldsymbol{\alpha}=2\pi(\sqrt{3}\hat{x}+\hat{z})/(3d)$ and
$\boldsymbol{\beta}=4\pi/(3d)\hat{z}$.  The Wigner-Seitz unit cell and the
first Brillouin zone are thus the parallelograms formed, respectively,
by the pairs of vectors $(\mathbf{a}$, $\mathbf{b})$ and
$(\boldsymbol{\alpha}$, $\boldsymbol{\beta})$.  Both may equivalently be taken
as being centred on a point of sixfold rotational symmetry in their
respective spaces.  Thus, the Wigner-Seitz unit cell may be taken as
being bounded by the sides of a primitive hexagon of side length $d$
as in Fig.\ \ref{model_pattern}.  In this case the first Brillouin
zone is also a hexagon, now of side length $4\pi/(3\sqrt{3}d)$, and which is
also rotated by 90$^{\circ}$ with respect to the Wigner-Seitz hexagon.

The classical ($s \rightarrow \infty$) version of the $J_{1}$--$J_{2}$--$J_{3}$ model of Eq.\ (\ref{Hamiltonian}) on the honeycomb lattice already itself displays a rich $T=0$ GS phase diagram (see, e.g., Refs.\ \cite{Rastelli:1979_honey,Fouet:2001_honey,Mulder:2010_honey}).  The generic stable GS phase is a coplanar spiral configuration of spins defined  
by a wave vector {\bf Q}, together with an angle $\theta$ that is the
relative orientation of the two spins in the same unit cell $i$ characterized by the lattice
vector $\mathbf{R}_{i}$.  The two classical spins in unit
cell $i$ are given by
\begin{equation}
\mathbf{s}_{i,\rho}=-s[\cos(\mathbf{Q}\cdot\mathbf{R}_{i}+\theta_{\rho})\hat{z}_{s}+\sin(\mathbf{Q}\cdot\mathbf{R}_{i}+\theta_{\rho})\hat{x}_{s}]\,; \quad \rho={\cal A},\,{\cal B}\,,  \label{eq_classical-spins}
\end{equation}
where $\hat{x}_{s}$ and $\hat{z}_{s}$ are two orthogonal unit vectors
that define the spin-space plane, as shown in Fig.\
\ref{model_pattern}.  We choose the two angles $\theta_{\rho}$ such that
$\theta_{{\cal A}}=0$ and $\theta_{{\cal B}}=\theta$.

When all three bonds are AFM in nature (i.e., $J_{m}>0$; $m=1,2,3$), as considered here, it has been shown \cite{Fouet:2001_honey,Mulder:2010_honey} that the
classical model has a $T=0$ GS phase diagram consisting of three
different phases.  With reference to an origin at the centre of the hexagonal Wigner-Seitz cell, one may show that one value
of the spiral wave vector $\mathbf{Q}$ that minimizes the classical GS energy of the model is given by
\begin{equation}
\mathbf{Q}=\frac{2}{\sqrt{3}d}\cos^{-1}\left[\frac{(J_{1}-2J_{2})}{4(J_{2}-J_{3})}\right]\hat{x}\,,  \label{eq_spiral-wave-factor}
\end{equation}
together with $\theta=\pi$.  
Equation (\ref{eq_spiral-wave-factor}) is clearly only valid when
\begin{equation}
-1 \leq \frac{J_{1}-2J_{2}}{4(J_{2}-J_{3})} \leq 1 \,. \label{eq_conditon_Q}
\end{equation}
If we define $x \equiv J_{2}/J_{1}$ and $y \equiv J_{3}/J_{1}$, Eq.\ (\ref{eq_conditon_Q}) is equivalent to the inequalities,
\begin{equation}
y \leq \frac{3}{2}x-\frac{1}{4}\,; \quad y \leq \frac{1}{2}x+\frac{1}{4}\,.  \label{eq_xy}
\end{equation}
In the positive quadrant (i.e., $x \geq 0$, $y \geq 0$) of the $xy$
plane the classical model has the spiral phase described by the wave
vector $\mathbf{Q}$ of Eq.\ (\ref{eq_spiral-wave-factor}) and $\theta=\pi$
as the stable GS phase in the region defined by Eq.\ (\ref{eq_xy}).

Everywhere on the boundary line $y = \frac{3}{2}x-\frac{1}{4}$ of the
spiral phase, $\mathbf{Q} = \mathbf{\Gamma} = (0,0)$, which simply
describes the N\'{e}el AFM phase shown in Fig.\
\ref{model_pattern}(a).  Similarly, everywhere on the other boundary
line $y = \frac{1}{2}x+\frac{1}{4}$,
$\mathbf{Q}=2\pi/(\sqrt{3}d)\hat{x}$, which describes the collinear
striped AFM phase shown in Fig.\ \ref{model_pattern}(b).  Both the
phase transitions between the spiral and N\'{e}el phases and between
the spiral and striped phases are clearly continuous ones.  The two
phase boundaries meet at the tricritical point
$(x,y)=(\frac{1}{2},\frac{1}{2})$.  Finally, one can also easily show
that there is a first-order phase transition between the two collinear
AFM phases along the line $x=\frac{1}{2}$, $y>\frac{1}{2}$.  To
summarize, in the regime where $J_{1}>0$ and $x \geq 0$, $y \geq 0$,
the classical version of the honeycomb-lattice
$J_{1}$--$J_{2}$--$J_{3}$ model has three stable GS phases at $T=0$.
These are: (a) a spiral phase for $0<y<\frac{3}{2}x-\frac{1}{4}$,
$\frac{1}{6}<x<\frac{1}{2}$ and $0<y<\frac{1}{2}x+\frac{1}{4}$,
$x>\frac{1}{2}$; (b) a N\'{e}el AFM phase for $y>0$, $0<x<\frac{1}{6}$
and $y>\frac{3}{2}x-\frac{1}{4}$, $\frac{1}{6}<x<\frac{1}{2}$; and (c)
a striped collinear AFM phase for $y>\frac{1}{2}x+\frac{1}{4}$,
$x>\frac{1}{2}$.  Clearly, along the line $y=x$ considered here, which
includes the tricritical point at ($\frac{1}{2},\frac{1}{2}$), there
are just two stable GS phases, namely the collinear N\'{e}el and
striped AFM phases.

It is worth noting that both the spiral and the striped states described
by the wave vector of Eq.\ (\ref{eq_spiral-wave-factor})
and its appropriate limiting form (and $\theta=\pi$) have two other equivalent states
rotated by $\pm \frac{2}{3}\pi$ in the honeycomb $xz$ plane.  We also
note that in the limiting case when the $J_{2}$ bond dominates (i.e.,
when $x \rightarrow \infty$ for a fixed finite value of $y$), the
spiral pitch angle
$\phi=\cos^{-1}[\frac{1}{4}(J_{1}-2J_{2})/(J_{2}-J_{3})] \rightarrow
\frac{2}{3}\pi$.  Clearly, in this limit, the classical model reduces
to two disconnected Heisenberg antiferromagnets (HAFs) on
interpenetrating triangular lattices, each with the 3-sublattice
N\'{e}el ordering of NN spins (on each triangular lattice) oriented
at an angle $\frac{2}{3}\pi$ to one another.  Precisely in this limit
the wave vector $\mathbf{Q}$ of Eq.\ (\ref{eq_spiral-wave-factor})
becomes one of the six corners,
$\mathbf{K}^{(1)}=4\pi/(3\sqrt{3}d)\hat{x}$, of the hexagonal first
Brillouin zone.  The two inequivalent corner vectors thus describe the
two distinct 3-sublattice N\'{e}el orderings for a classical
triangular-lattice HAF.

For spiral pitch angles in the range $\frac{2}{3}\pi < \phi
\leq \pi$ the wave vector $\mathbf{Q}$ of Eq.\
(\ref{eq_spiral-wave-factor}) lies outside the first hexagonal Brillouin
zone.  It can equivalently be mapped back inside this range of values,
when $\mathbf{Q}$ then moves continuously from a corner at position
$\mathbf{K}^{(3)}=2\pi(-\hat{x}+\sqrt{3}\hat{z})/(3\sqrt{3}d)$ along
an edge to its midpoint at $\mathbf{M}^{(2)}=2\pi/(3d)\hat{z}$.  Thus,
the striped AFM state shown in Fig.\ \ref{model_pattern}(b) may
equivalently be described by the ordering wave vector
$\mathbf{Q}=\mathbf{M}^{(2)}$ (with the relative angle between the two
triangular sublattices ${\cal A}$ and ${\cal B}$ being $\theta=\pi$).
The two other equivalent striped states have wave vectors
corresponding to the other two inequivalent midpoints of the hexagonal
Brillouin zone edges, at
$\mathbf{M}^{(1)}=\pi(\sqrt{3}\hat{x}+\hat{z})/(3d)$ and
$\mathbf{M}^{(3)}=\pi(-\sqrt{3}\hat{x}+\hat{z})/(3d)$ (with
$\theta=0$ in these two cases).

While Eq.\ (\ref{eq_classical-spins}) is generally sufficient to
describe the classical GS spin configuration
\cite{Villain:1977_ordByDisord}, it relies on the assumption that the
GS order either is unique (up to a global rotation of all spins by the
same amount) or exhibits, at most, a discrete degeneracy (e.g., as
associated with rotations of the wave vector $\mathbf{Q}$ by $\pm
120^{\circ}$ about the $\hat{y}$ axis).  Nevertheless, the assumption can be shown to be false
for special values of $\mathbf{Q}$
\cite{Fouet:2001_honey,Villain:1977_ordByDisord}, which include the
cases when $\mathbf{Q}$ equals either one half or one quarter of a
reciprocal lattice vector $\mathbf{G}$.  This includes precisely the
case for the striped states for which the wave vectors
$\mathbf{Q}=\mathbf{M}^{(i)}$, $i=1,2,3$, equal one half of
corresponding reciprocal lattice vectors.  In this case, it has been
shown \cite{Fouet:2001_honey} that the GS ordering now
spans a 2D manifold of non-planar spin configurations, all of which are
degenerate in energy with those of the striped states.

Classical spin-lattice systems that display such
an infinitely degenerate family (IDF) of GS phases in some region of
their $T=0$ phase space are well known to be prime candidates for the
emergence of novel quantum phases with no classical counterparts in the
corresponding quantum systems.  Quantum fluctuations then often act to
lift this accidental GS degeneracy (either wholly or in part) by the
order by disorder mechanism
\cite{Villain:1977_ordByDisord,Villain:1980_ordByDisord,Shender:1982_ordByDisord}
in favour of just one (or several) member(s) of the classical IDF.  As
we noted previously in Sec.\ \ref{intro}, the striped collinear state
is indeed energetically selected by quantum fluctuations
\cite{Fouet:2001_honey} in the present case in the large-$s$ limit
\cite{Rastelli:1979_honey} where first-order linear spin-wave theory
(LSWT) becomes exact.

Of course, quantum fluctuations can also be expected in such cases of
macroscopic degeneracy of the classical GS phase, to destroy
completely the magnetic LRO, as we discussed in Sec.\ \ref{intro},
Clearly, this is most likely to occur for small values of $s$, when
the results of LSWT become less likely to remain valid and when
quantum fluctuations become larger.  Since the specific case when
$J_{3}=J_{2}\equiv \kappa J_{1}$ includes the classical tricritical point at
$\kappa_{{\rm cl}} = \frac{1}{2}$, which is the point of
maximum classical frustration, we restrict further attention to this
potentially rich regime in the entire parameter space of the $J_{1}$--$J_{2}$--$J_{3}$ model.
Along this line, the classical ($s \rightarrow \infty$) model at $T=0$
undergoes a first-order transition from the N\'{e}el phase, which is
the stable GS phase for $\kappa < \kappa_{{\rm cl}}$, to the striped
phase which is the stable GS phase for $\kappa > \kappa_{{\rm cl}}$.  From our above discussion it
is clear that the most promising regime for novel quantum states, with
no classical counterparts, to emerge is the region around $\kappa
\approx \frac{1}{2}$.

In an earlier paper \cite{DJJF:2011_honeycomb} the
$J_{1}$--$J_{2}$--$J_{3}$ model with $J_{3}=J_{2}\equiv \kappa J_{1} >
0$ was studied for the case $s=\frac{1}{2}$ in the window $0 \leq
\kappa \leq 1$, using the CCM.  It was found
\cite{DJJF:2011_honeycomb} that the classical ($s \rightarrow \infty$)
transition at $\kappa_{{\rm cl}}=\frac{1}{2}$ is split in the
$s=\frac{1}{2}$ case into two transitions at $\kappa_{c_{1}} < \kappa_{{\rm
    cl}}$ and $\kappa_{c_{2}} > \kappa_{{\rm cl}}$, with the N\'{e}el phase
surviving for $\kappa < \kappa_{c_{1}} \approx 0.47$ and the striped
phase for $\kappa > \kappa_{c_{2}} \approx 0.60$.  A paramagnetic
phase, with no discernible magnetic LRO, was indeed found to exist in
the intermediate regime $\kappa_{c_{1}} < \kappa < \kappa_{c_{2}}$.
CCM calculations were also performed to measure the susceptibilities of
the two AFM phases on either side of the paramagnetic regime against
the formation of plaquette valence-bond crystalline (PVBC) order
\cite{DJJF:2011_honeycomb}.  It was thereby concluded that the paramagnetic
state was most likely one with PVBC order over the entire intermediate
regime $\kappa_{c_{1}} < \kappa < \kappa_{c_{2}}$.  On the basis of
all the CCM calculations (i.e., for the GS energy per spin $E/N$, the
GS magnetic order parameter $M$, and the susceptibility $\chi_{p}$
against the formation of PVBC order, for the two AFM states on either
side of the intermediate regime), the accumulated evidence pointed
towards the quantum phase transition (QPT) at the quantum critical
point (QCP) $\kappa = \kappa_{c_{2}}$ being a first-order one, just as is
the classical phase transition at $\kappa = \kappa_{{\rm cl}}$.  By
contrast, the QPT at $\kappa = \kappa_{c_{1}}$ appeared to be a
continuous one on the basis of the CCM results presented.  Since the
quasiclassical N\'{e}el phase and the quantum PVBC phase break different
symmetries, however, the usual Landau-Ginzburg-Wilson scenario of continuous
phase transitions is inapplicable, and it was suggested that the CCM
results \cite{DJJF:2011_honeycomb} provide strong evidence for the QPT
at $\kappa=\kappa_{c_{1}}$ being of the deconfined quantum critical
type \cite{Senthil:2004_Science_deconfinedQC,Senthil:2004_PRB_deconfinedQC}.

In view of the qualitative differences between the GS phase diagrams
of the above spin-$\frac{1}{2}$ and classical ($s \rightarrow \infty$)
versions of the model, it is obviously of considerable interest to
examine the model in the case where the spin quantum number $s >
\frac{1}{2}$.  One of the great strengths of the CCM is that it is
relatively straightforward, both conceptually and computationally, to
examine a given spin-lattice model for different values of $s$, within
a unified and consistent hierarchy of approximations.  Hence, we now
use the method to examine the $J_{1}$--$J_{2}$--$J_{3}$ model on the
honeycomb lattice, along the line $J_{3}=J_{2}\equiv \kappa J_{1} >
0$, with $0 \leq \kappa \leq 1$, for the cases
$s=1,\,\frac{3}{2},\,2\,,\frac{5}{2}$, in order to compare them both
with the extreme quantum limit ($s=\frac{1}{2}$) case and with the
classical ($s \rightarrow \infty$) case.

\section{The coupled cluster method}
\label{ccm_section}
We briefly describe here the key features of the CCM, and refer the
interested reader to the extensive literature (and see, e.g., Refs.\
\cite{Bishop:1978_ccm,Bishop:1982_ccm,Arponen:1983_ccm,Bishop:1987_ccm,Arponen:1991_ccm,Bishop:1991_TheorChimActa_QMBT,Bishop:1998_QMBT_coll,Fa:2004_QM-coll}
and references cited therein) for further details.  To implement the
method in practice one first needs to choose a suitable normalized
model (or reference) state $|\Phi\rangle$, against which the
correlations present in the exact GS wave function can be
incorporated.  The properties required of $|\Phi\rangle$ are described
more fully below, but in general terms it plays the role of a
generalized vacuum state.  For the present study suitable choices for
the model state $|\Phi\rangle$ will turn out to be the two
quasiclassical AFM states (viz., the N\'{e}el and collinear striped
states) that form the stable GS phases of the classical version of the
model under consideration in their respective regimes of the $T=0$
phase diagram.

The exact GS ket- and bra-state wave functions, $|\Psi\rangle$ and
$\langle\tilde{\Psi}|$, respectively, are chosen to have the normalization
conditions,
\begin{equation}
\langle\tilde{\Psi}|\Psi\rangle = \langle{\Phi}|\Psi\rangle =
\langle{\Phi}|\Phi\rangle \equiv 1\,.   \label{norm_conditions}
\end{equation}
These exact states are now parametrized with respect to the model
state $|\Phi\rangle$ in the exponentiated forms,
\begin{equation}
|\Psi\rangle=e^{S}|\Phi\rangle\,; \quad \langle\tilde{\Psi}|=\langle\Phi|\tilde{S}e^{-S}\,,  \label{exp_para}
\end{equation}
that are a characteristic
hallmark of the CCM.  Although the correlation operator $\tilde{S}$ may formally be expressed in terms of its counterpart $S$ as
\begin{equation}
\langle\Phi|\tilde{S} = \frac{\langle\Phi|e^{S^{\dagger}}e^{S}}{\langle\Phi|e^{S^{\dagger}}e^{S}|\Phi\rangle}\,,  \label{correlation-opererators-relationship}
\end{equation}
by using Hermiticity, the CCM chooses not to restrain this relationship between $|\Psi\rangle$ and $\langle\tilde{\Psi}|$ explicitly.  Instead, the 
two correlation operators $S$ and $\tilde{S}$ are 
formally decomposed independently as
\begin{equation}
S=\sum_{I\neq 0}{\cal S}_{I}C^{+}_{I}\,; \quad \tilde{S}=1+\sum_{I\neq 0}\tilde{{\cal S}}_{I}C^{-}_{I}\,,  \label{correlation_oper}
\end{equation}
where $C^{+}_{0}\equiv 1$ is defined to be the identity operator in the many-body Hilbert space, and
where the set index $I$ denotes a complete set of single-particle
configurations for all $N$ particles.  What is required of $|\Phi\rangle$ and the set of (multiconfigurational) creation operators $\{C^{+}_{I}\}$ is that $|\Phi\rangle$ is a fiducial (or cyclic) vector with respect to these operators, i.e., as a generalized vacuum state.  Explicitly we require that the set of states $\{C^{+}_{I}|\Phi\rangle\}$ form a complete basis for the ket-state Hilbert space, and that
\begin{equation}
\langle\Phi|C^{+}_{I} = 0 = C^{-}_{I}|\Phi\rangle\,, \quad \forall I
\neq 0\,,  \label{creat-destruct-operators-relationship}
\end{equation}
where the destruction operators $C^{-}_{I} \equiv
(C^{+}_{I})^{\dagger}$.  Lastly, and importantly, we require that all
members of the complete set of operators $\{C^{+}_{I}\}$ are mutually
commuting.

The rather general CCM paramerizations of Eqs.\
(\ref{exp_para})--(\ref{creat-destruct-operators-relationship})
have several immediate important consequences.  While Hermiticity is
not made explicit, and while the exact correlation operators $S$ and
$\tilde{S}$ will certainly fulfill Eq.\
(\ref{correlation-opererators-relationship}), when approximations are made
(e.g., by truncating the sums over configuration $I$ in Eq.\
(\ref{correlation_oper}), as is usually done in practice),
Hermiticity may be only approximately maintained.  Against this loss,
however, come several advantages, which usually far outweigh it.
Firstly, the CCM parametrizations guarantee that the Goldstone linked
cluster theorem is exactly preserved, as we describe in more detail
below, even if truncations are made in Eq.\ (\ref{correlation_oper}).
In turn, this feature guarantees size-extensivity at any such level of
truncation, so that the GS energy, for example, is always calculated as
an extensive variable.  Thus, the CCM has the first advantage that we
may work from the outset in the thermodynamic limit ($N \rightarrow
\infty$), thereby obviating the need for any finite-size scaling, as
is required in most alternative methods.  A second key feature of the
CCM, which is guaranteed by its exponentiated parametrizations, is
that it also exactly preserves the important Hellmann-Feynman theorem
at any level of truncation or approximate implementation.

Clearly, a knowledge of the CCM $c$-number correlation
coefficients $\{{\cal S}_{I}, \tilde{{\cal S}}_{I}\}$ completely suffices to determine the GS expectation value of any operator.  They are now found by minimization of the GS energy expectation functional,
\begin{equation}
\bar{H}=\bar{H}({\cal S}_{I},{\tilde{\cal S}_{I}}) \equiv
\langle\Phi|\tilde{S}e^{-S}He^{S}|\Phi\rangle\,,  \label{eq_GS_E_xpect_funct}
\end{equation}
from Eq.\ (\ref{exp_para}),
with respect to each of the coefficients $\{{\cal S}_{I},{\tilde{\cal
    S}}_{I}\,; \forall I \neq 0\}$ separately.  Variation of $\bar{H}$ from Eq.\ (\ref{eq_GS_E_xpect_funct}), with
respect to ${\tilde{\cal S}}_{I}$ from Eq.\ (\ref{correlation_oper}), immediately yields 
\begin{equation}
\langle\Phi|C^{-}_{I}e^{-S}He^{S}|\Phi\rangle = 0\,, \quad \forall I \neq 0\,,  \label{ket_eq}
\end{equation}
which is a coupled set of {\it non-linear} equations for the
coefficients $\{{\cal S}_{I}\}$, with the same number of equations as
parameters.  A similar variation of $\bar{H}$ from Eq.\
(\ref{eq_GS_E_xpect_funct}), with respect to ${\cal S}_{I}$ from Eq.\ (\ref{correlation_oper}) yields
\begin{equation}
\langle\Phi|\tilde{S}e^{-S}[H,C^{+}_{I}]e^{S}|\Phi\rangle=0\,, \quad \forall I \neq 0\,,  \label{bra_eq}
\end{equation}
as a coupled set of {\it linear} equations for the coefficients $\{{\tilde{\cal S}}_{I}\}$,
again with the same number of equations as parameters, once the
coefficients $\{{\cal S}_{I}\}$ are used as input after Eq.\
(\ref{ket_eq}) has been solved for them.

The GS energy $E$, which is simply the value of $\bar{H}$ from Eq.\ (\ref{eq_GS_E_xpect_funct}) at the minimum, may then be expressed as
\begin{equation}
E=\langle\Phi|e^{-S}He^{S}|\Phi\rangle=\langle\Phi|He^{S}|\Phi\rangle\,, \label{eq_GS_E}
\end{equation}
using Eqs.\ (\ref{ket_eq}) and (\ref{bra_eq}).  By making use of Eq.\
(\ref{eq_GS_E}), we may rewrite the set of linear equations (\ref{bra_eq}) in the equivalent form,
\begin{equation}
\langle\Phi|\tilde{S}(e^{-S}He^{S}-E)C^{+}_{I}|\Phi\rangle=0\,, \quad \forall I \neq 0\,,  \label{bra_eq_alternative}
\end{equation}
which is just a set of generalized linear eigenvalue equations for the
set of coefficients $\{\tilde{\cal S}_{I}\}$.

Up to this point in the CCM procedure and implementation we have made no approximations.  However, clearly
Eqs.\ (\ref{ket_eq}) that determine the creation coefficients $\{{\cal S}_{I}\}$ are
intrinsically highly nonlinear in view of the exponential terms.  Hence one may ask if we now need to make truncations to evaluate these terms.  We note, though, that these always
appear in the equations to be solved in the combination $e^{-S}He^{S}$ of a similarity transformation of the Hamiltonian.  This may itself be expanded as the well-known nested commutator sum,
\begin{equation}
e^{-S}He^{S} = \sum^{\infty}_{n=0}\frac{1}{n!}[H,S]_{n}\,,  \label{eq_expon_nested_commutator}
\end{equation}
where $[H,S]_{n}$ is an $n$-fold nested commutator, defined
iteratively as
\begin{equation}
[H,S]_{n}=[[H,S]_{n-1},S]\,; \quad [H,S]_{0}=H\,.
\end{equation}
Another key feature of the CCM is that this otherwise infinite sum now
(usually) terminates exactly at some finite order, when used in the
equations to be solved, due to the facts that all of the terms in the
expansion of Eq.\ (\ref{correlation_oper}) for $S$ commute with one
another and that $H$ itself is (usually, as here) of finite order in the
relevant single-particle operators.  For example, if $H$ contains up
to $m$-body interactions, in its second-quantized form it contains
sums of terms involving products of up to $2m$ single-particle
(destruction and creation) operators, and the sum in Eq.\
(\ref{eq_expon_nested_commutator}) will terminate exactly with the
term $n=2m$.  In our present case where the Hamiltonian of Eq.\
(\ref{Hamiltonian}) is bilinear in the SU(2) spin operators, the sum
terminates at $n=2$.  Finally, we also note here that the fact that
all of the operators in the set $\{C^{+}_{I}\}$ that comprise $S$ by
Eq.\ (\ref{correlation_oper}) commute with one another, automatically
guarantees that all (nonzero) terms in the sum of Eq.\
(\ref{eq_expon_nested_commutator}) are linked to the Hamiltonian.
Unlinked terms simply cannot appear, and hence the Goldstone theorem
and size-extensivity are satisfied, at any level of truncation.

Hence, for any implementation of the CCM, the {\it only} approximation
made in practice is to restrict the set of multiconfigurational
set-indices $\{I\}$ that are retained in the expansions of Eq.\
(\ref{correlation_oper}) for the correlation operators
$\{S,\tilde{S}\}$ to some appropriate (finite or infinite) subset.
How this choice is made must clearly depend on the problem at hand and
on the particular choices that have been made for the model state
$|\Phi\rangle$ and the associated set of operators $\{C^{+}_{I}\}$.  Let
us, therefore, now turn to how such choices are made for the present
model in particular and for quantum spin-lattice models in general.

The simplest choice of model state $|\Phi\rangle$ for a quantum
spin-lattice problem is a straightforward independent-spin product
state in which the spin projection (along some specified quantization axis) of the spin on each lattice site is
specified independently.  The two quasiclassical collinear AFM states
shown in Figs.\ \ref{model_pattern}(a) and \ref{model_pattern}(b),
viz., the N\'{e}el and striped states, are examples.  In order to
treat all such states in the same way it is very convenient to make a
passive rotation of each spin independently (i.e., by making a
suitable choice of local spin quantization axes on each site independently), so that
on every site the spin points downwards, say, in the negative $z_{s}$
direction, as in the spin-coordinate frame shown in Fig.\
\ref{model_pattern}.  Such rotations are clearly just unitary
transformations that leave the basic SU(2) spin commutation relations
unchanged.  In this way each lattice site $k$ is completely equivalent
to all others, and all such independent-spin product model states now
take the universal form
$|\Phi\rangle=|$$\downarrow\downarrow\downarrow\cdots\downarrow\rangle$.

In this representation it is now clear that $|\Phi\rangle$ can indeed
be regarded as a fiducial vector with respect to a set of mutually
commuting creation operators $\{C^{+}_{I}\}$, which may now be chosen
as a product of single-spin raising operators, $s^{+}_{k} \equiv
s^{x}_{k}+is^{y}_{k}$, such that $C^{+}_{I} \rightarrow
s^{+}_{k_{1}}s^{+}_{k_{2}}\cdots s^{+}_{k_{n}};\; n=1,2,\cdots , 2sN$.
The corresponding set index $I$ thus becomes a set of lattice-site
indices, $I \rightarrow \{k_{1},k_{2},\cdots , k_{n};\; n=1,2,\cdots ,
2sN\}$, in which each site index may be repeated up to $2s$ times.
Once the local spin coordinates have been selected by the above procedure
(i.e., for the given model state $|\Phi\rangle$), one simply
re-expresses the Hamiltonian $H$ in terms of them.

We now turn to the choice of approximation scheme, which hence simply
involves a choice of which configurations $\{I\}$ to retain in the
decompositions of Eq.\ (\ref{correlation_oper}) for the CCM
correlation operators $(S,\tilde{S})$.  A powerful and rather general
such scheme, the so-called SUB$n$--$m$ scheme, retains the
configurations involving a maximum of $n$ spin-flips (where each
spin-flip requires the action of a spin-raising operator $s^{+}_{k}$
acting once) spanning a range of no more than $m$ contiguous sites on
the lattice.  A set of lattice sites is defined to be contiguous if
every site in the set is the NN of at least one other in the set (in a
specified geometry).  Clearly, as both indices become indefinitely
large, the approximation becomes exact.  Different schemes can be
defined according to how each index approaches infinity.

For example, if we first let $m \rightarrow \infty$, we arrive at the
so-called SUB$n$ $\equiv$ SUB$n$--$\infty$ scheme, which is the
approximation scheme most commonly employed, more generally, for
systems defined in a spatial continuum, such as atoms and molecules in
quantum chemistry \cite{Bartlett:1989_ccm} or finite atomic nuclei or
nuclear matter in nuclear physics \cite{Kummel:1978_ccm} (and see,
e.g., Refs.\ \cite{Bishop:1978_ccm,Bishop:1982_ccm} for further
details).  By contrast to continuum theories, for which the notion of
contiguity is not easily applicable, in lattice theories both indices
$n$ and $m$ may be kept finite.  A very commonly used scheme is the
so-called LSUB$m$ scheme
\cite{Fa:2004_QM-coll,Zeng:1998_SqLatt_TrianLatt}, defined to retain,
at the $m$th level of approximation, all spin clusters described by
multispin configurations in the index set $\{I\}$ defined over any
possible lattice animal (or polyomino) of size $m$ on the lattice.
Again, such a lattice animal is defined in the usual graph-theoretic
sense to be a configured set of contiguous (in the above sense) sites
on the lattice.  Clearly, the LSUB$m$ scheme is equivalent to the
SUB$n$--$m$ scheme when $n=2sm$ for particles of spin quantum number
$s$, i.e., LSUB$m \equiv$ SUB$2sm$--$m$.  The LSUB$m$ scheme was precisely the
truncation scheme used in our previous study of the present model for
the case $s=\frac{1}{2}$ \cite{DJJF:2011_honeycomb}.

At a given $m$th level of LSUB$m$ approximation the number $N_{f}$ of
fundamental spin configurations that are distinct (under the symmetries
of the lattice and the specified model state), which are retained is
lowest for $s=\frac{1}{2}$ and rises sharply as $s$ is increased.
Since $N_{f}$ typically also increases rapidly (typically, faster than
exponentially) with the truncation index $m$, an alternative scheme
for use in cases $s > \frac{1}{2}$ is to be set $m=n$ and hence employ
the resulting SUB$n$--$n$ scheme.  Clearly, the two schemes are
equivalent only for the case $s=\frac{1}{2}$, for which LSUB$m$
$\equiv$ SUB$m$--$m$.  We note too that the numbers $N_{f}$ of
fundamental configurations at a given SUB$n$--$n$ level are still
higher for the cases $s>\frac{1}{2}$ considered here than for the case
$s=\frac{1}{2}$.  Thus, whereas for the the present model we were able
to perform LSUB$m$ calculations with $m \leq 12$ previously for the
case $s=\frac{1}{2}$ \cite{DJJF:2011_honeycomb}, we are now restricted
for the cases $s>\frac{1}{2}$ considered here to perform SUB$n$--$n$
calculations with $n \leq 10$, with similar amounts of supercomputer
resources available.  Thus, for example, for the case $s=\frac{1}{2}$,
at the LSUB12 level of approximation we have $N_{f}=103,097 (250,891)$
using the N\'{e}el (striped) state as the CCM model state.  By
comparison, at the SUB10--10 level of approximation we have
$N_{f}=219,521 (552,678)$ for the case $s=1$, and $N_{f}=538,570
(1,436,958)$ for the case $s=\frac{5}{2}$, in each case using the
N\'{e}el (striped) state as the CCM model state.  Just as before
\cite{DJJF:2011_honeycomb} we employ massively parallel computing
\cite{ccm_code} both to derive (with computer algebra) and to solve
(and see, e.g., Ref.\cite{Zeng:1998_SqLatt_TrianLatt}) the respective
coupled sets of CCM equations (\ref{ket_eq}) and
(\ref{bra_eq_alternative}).

Once the coefficients $\{{\cal S}_{I},{\tilde{{\cal S}}}_{I}\}$ retained in a given
SUB$n$--$n$ approximation have been calculated by solving Eqs.\
(\ref{ket_eq}) and (\ref{bra_eq_alternative}), we may calculate any GS
quantity at the same level of approximation.  Thus, for example, the
GS energy $E$ may be calculated from Eq.\ (\ref{eq_GS_E}) in terms
of the ket-state coefficients $\{{\cal S}_{I}\}$ alone.  Any other GS
quantity requires a knowledge also of the bra-state coefficients
$\{\tilde{{\cal S}}_{I}\}$.  For example, we also calculate here the
magnetic order parameter $M$, which is defined to be the average
on-site GS magnetization,
\begin{equation}
M = -\frac{1}{N}\sum^{N}_{k=1}\langle\Phi|\tilde{S}
  e^{-S}s^{z}_{k}e^{S}|\Phi\rangle\,,   \label{M_eq}
\end{equation}
in terms of the local rotated spin-coordinate frames that we have
described above.

As a last step, and as essentially the only approximation made in the
whole CCM implementation, we need to extrapolate the raw SUB$n$--$n$
data points for $E$ and $M$ to the exact $n \rightarrow \infty$ limit.
Although no exact extrapolation rules are known, a great deal of experience
has by now been accumulated for doing so, from the many applications
of the technique to a wide variety of spin-lattice problems that have
been examined with the method.  For the GS energy per spin, for
example, a very well tested and highly accurate extrapolation ansatz
(and see, e.g., Refs.\
\cite{DJJF:2011_honeycomb,PHYLi:2012_honeycomb_J1neg,Bishop:2012_honeyJ1-J2,Bishop:2012_honey_phase,Li:2012_honey_full,RFB:2013_hcomb_SDVBC,Fa:2004_QM-coll,Bishop:2000_XXZ,Kruger:2000_JJprime,Fa:2001_SqLatt_s1,Darradi:2005_Shastry-Sutherland,Darradi:2008_J1J2mod,Bi:2008_EPL_J1J1primeJ2_s1,Bi:2008_JPCM_J1xxzJ2xxz_s1,Bi:2009_SqTriangle,Bishop:2010_UJack,Bishop:2010_KagomeSq,Bishop:2011_UJack_GrtSpins,PHYLi:2012_SqTriangle_grtSpins,Li:2012_anisotropic_kagomeSq})
is
\begin{equation}
\frac{E(n)}{N} = a_{0}+a_{1}n^{-2}+a_{2}n^{-4}\,,     \label{extrapo_E}
\end{equation}
while for the magnetic order parameter $M$ different schemes have been
used in different situations.  Unsurprisingly, the GS expectation
values of other physical observables generally converge less rapidly
than the GS energy, i.e., with leading exponents less than two.  More
specifically, the leading exponent for $M$ tends to depend on the
amount of frustration present, generally being smaller for the most
highly frustrated cases.

Thus, for unfrustrated models or for models with only moderate amounts of
frustration present, a scaling ansatz for $M(n)$ with leading power $1/n$ (rather than $1/n^{2}$ as for the GS energy),
\begin{equation}
M(n) = b_{0}+b_{1}n^{-1}+b_{2}n^{-2}\,,   \label{M_extrapo_standard}
\end{equation}
has been found to work well in many cases (and see, e.g., Refs.\
\cite{PHYLi:2012_honeycomb_J1neg,Bishop:2012_honeyJ1-J2,Bishop:2012_honey_phase,RFB:2013_hcomb_SDVBC,Bishop:2000_XXZ,Kruger:2000_JJprime,Fa:2001_SqLatt_s1,Darradi:2005_Shastry-Sutherland,Bi:2009_SqTriangle,Bishop:2010_UJack,Bishop:2010_KagomeSq,Bishop:2011_UJack_GrtSpins,PHYLi:2012_SqTriangle_grtSpins}).
For systems that are either close to a QCP or for which the magnetic
order parameter $M$ for the phase under study is either zero or close
to zero, the extrapolation ansatz of Eq.\ (\ref{M_extrapo_standard})
tends to overestimate the extrapolated value and hence to predict a
somewhat too large value for the critical strength of the frustrating
interaction that is driving the respective phase transition.  In such
cases a great deal of evidence has now shown that a scaling ansatz
with leading power $1/n^{1/2}$ fits the SUB$n$--$n$ data much better.
Thus, as an alternative in those instances to Eq.\
(\ref{M_extrapo_standard}), a more appropriate scaling scheme (and see,
e.g., Refs.\
\cite{DJJF:2011_honeycomb,PHYLi:2012_honeycomb_J1neg,Bishop:2012_honeyJ1-J2,Bishop:2012_honey_phase,Li:2012_honey_full,RFB:2013_hcomb_SDVBC,Bi:2008_EPL_J1J1primeJ2_s1,Bi:2008_JPCM_J1xxzJ2xxz_s1,Li:2012_anisotropic_kagomeSq,Darradi:2008_J1J2mod})
is
\begin{equation}
M(n) = c_{0}+c_{1}n^{-1/2}+c_{2}n^{-3/2}\,.    \label{M_extrapo_frustrated}
\end{equation} 

Since the extrapolation schemes of Eqs.\
(\ref{extrapo_E})--(\ref{M_extrapo_frustrated}) contain three fitting
parameters, it is clearly preferable to use at least four SUB$n$--$n$
data points in each case.  Furthermore since the lowest-order SUB2--2
approximants are less likely to conform well to the extrapolation
schemes, we prefer to perform fits using SUB$n$--$n$ data with with $n
\geq 4$.

\section{Results}
\label{results_sec}
\begin{figure}
\begin{center}
\mbox{
\subfigure[]{\includegraphics[width=5.6cm,angle=270]{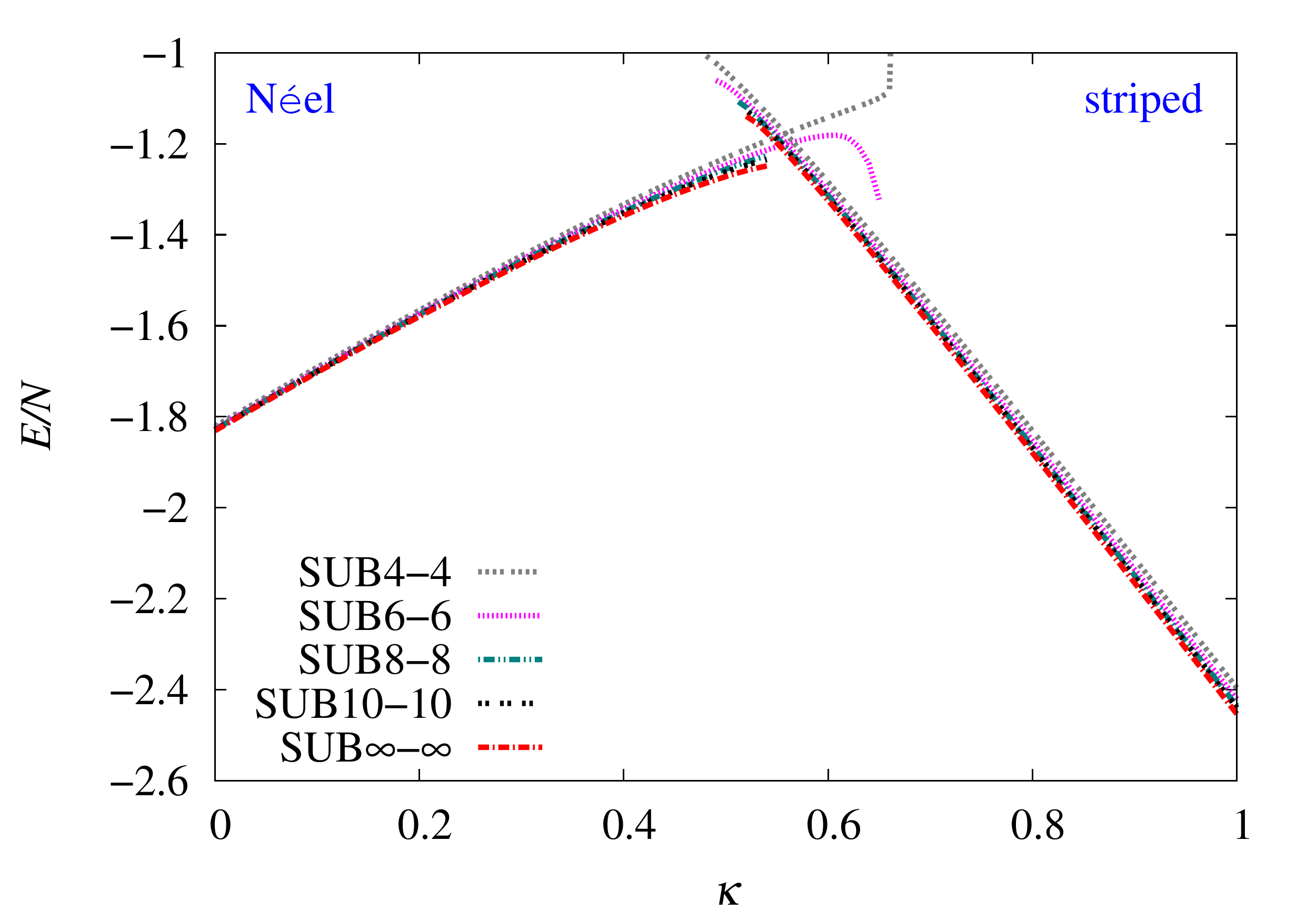}}
\subfigure[]{\includegraphics[width=5.6cm,angle=270]{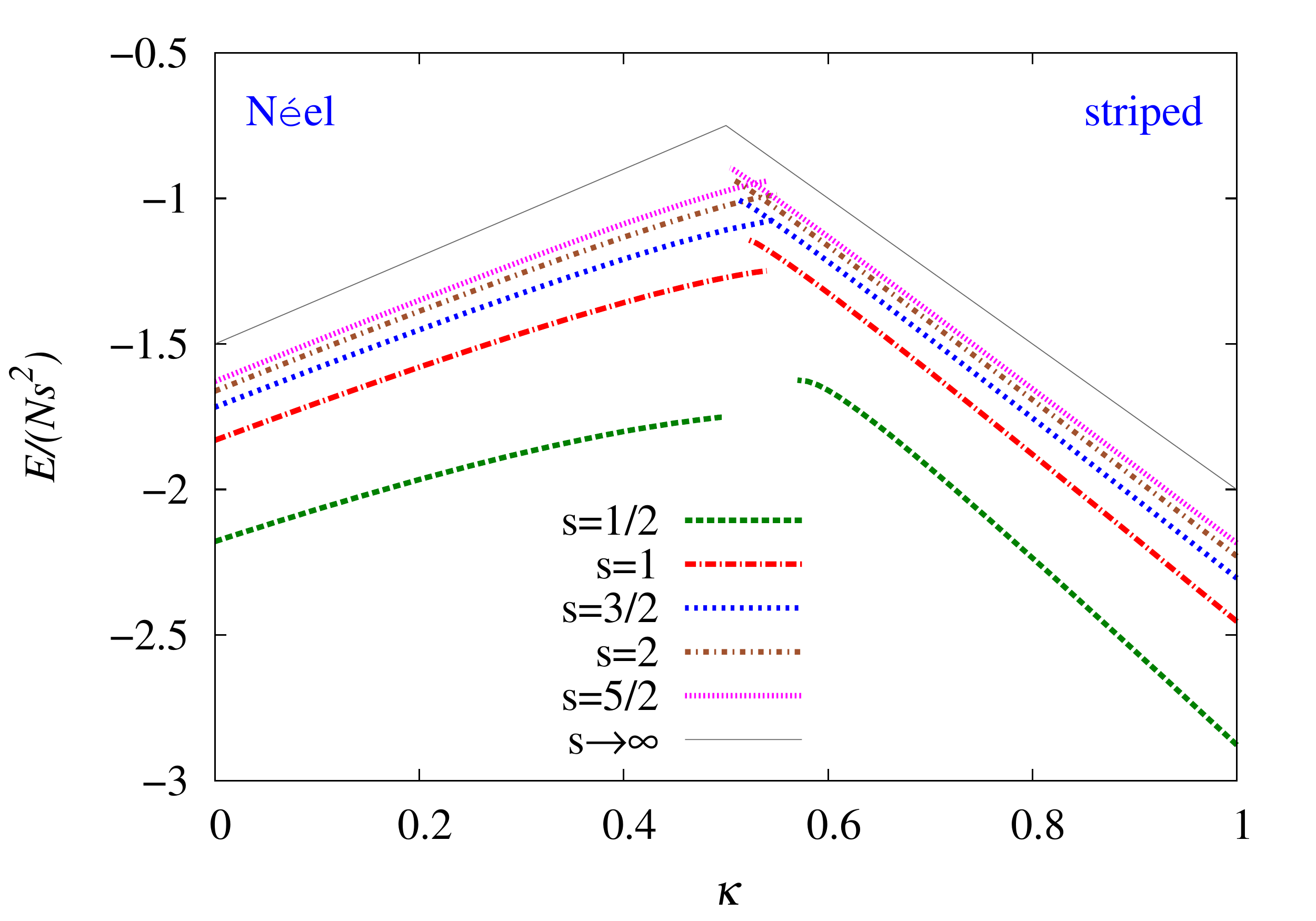}}
  }
\caption{CCM results for the GS energy per spin, $E/N$, for
  the $J_{1}$--$J_{2}$--$J_{3}$ model on the honeycomb lattice (with $J_{1} \equiv 1, J_{3}=J_{2} \equiv \kappa J_{1} > 0$), as a function of $\kappa$, 
  using the N\'{e}el and striped states as the CCM model states.  (a) The results for the $s=1$ model are shown using the SUB$n$--$n$
  approximations with $n=\{4,6,8,10\}$, together with the corresponding
  extrapolated SUB$\infty$--$\infty$ results obtained using equation (\ref{extrapo_E}), with this data set.  (b) We show extrapolated (SUB$\infty$--$\infty$) results for $E/(Ns^{2})$ as a function of $\kappa$, using equation
(\ref{extrapo_E}) together with the data set $n=\{6,8,10,12\}$ for the case $s=\frac{1}{2}$, and the corresponding data sets with $n=\{4,6,8,10\}$ for the four cases $s=1,\,\frac{3}{2},\,2\,,\frac{5}{2}$.  We also show the corresponding classical ($s \to \infty$) result.}
\label{E_multiSpins}
\end{center}
\end{figure}

We show in Fig.\ \ref{E_multiSpins}(a) our CCM results for the
GS energy per spin, $E/N$, of the spin-1 model at
various SUB$n$--$n$ levels of approximation with $n=\{4,6,8,10\}$,
using both the N\'{e}el and striped AFM states as separate choices of
the CCM model state.
We observe very clearly that the results converge very rapidly with
increasing values of the truncation index $n$, and we also show the
extrapolated $n \rightarrow \infty$ results, $a_{0}$, from Eq.\
(\ref{extrapo_E}).  We also observe that each of the energy curves
based on a particular model state terminates at a critical value of
the frustration parameter $\kappa$ that depends on the SUB$n$--$n$
approximation used.  Beyond those critical values no real solutions
can be found to the corresponding CCM equations (\ref{ket_eq}).  Such
termination points of the CCM coupled equations are very common in
practice, and are well understood (see, e.g., Refs.\
\cite{Bishop:2012_honeyJ1-J2,Fa:2004_QM-coll,Bi:2009_SqTriangle}).
They are direct manifestations of the corresponding QCP in the system,
at which the respective form of magnetic LRO in the model state used
melts.  As is usually the case, the CCM SUB$n$--$n$ solutions for a
given finite value of $n$ and for a given phase extend beyond the
actual SUB$\infty$--$\infty$ QCP, i.e., into the unphysical regime
beyond the termination point.  The extent of the unphysical regime
diminishes (to zero) as the truncation order $n$ increases (to the
exact $n \rightarrow \infty$ limit).

In Fig.\ \ref{E_multiSpins}(b) we compare the corresponding
extrapolated curves for the scaled GS energy per spin, $E/(Ns^{2})$,
using both the N\'{e}el and striped AFM states separately as our
choice of CCM model state, for the five cases
$s=\frac{1}{2},\,1,\,\frac{3}{2},\,2\,,\frac{5}{2}$.  In each case the
extrapolation is performed with Eq.\ (\ref{extrapo_E}).  For the case
$s=\frac{1}{2}$ alone the input SUB$n$--$n$ data points are
$n=\{6,8,10,12\}$, while for each of the cases $s>\frac{1}{2}$ the
input set is $n=\{4,6,8,10\}$.  We also show in Fig.\
\ref{E_multiSpins}(b) the corresponding classical ($s \rightarrow
\infty$) results, $E^{{\rm
    N\acute{e}el}}_{{\rm cl}}/(Ns^{2})=\frac{3}{2}(-1+\kappa)$ and $E^{{\rm
    striped}}_{{\rm cl}}/(Ns^{2})=\frac{1}{2}(1-5\kappa)$.  We observe clear
preliminary evidence from Fig.\ \ref{E_multiSpins}(b) for an
intermediate phase (between the phases with N\'{e}el and striped
magnetic LRO) in the $s=1$ case, although with a range of stability in
the frustration parameter $\kappa$ now markedly less than in the
$s=\frac{1}{2}$ case.  The preliminary evidence from the energy
results is also that there is no such intermediate phase present in
each of the cases $s > 1$.  Lastly, Fig.\ \ref{E_multiSpins}(b) also
shows that, at least so far as the energy results are concerned, all
cases with $s \gtrsim 2$ are rather close to the classical limit.

\begin{figure}
\begin{center}
\mbox{
\subfigure[]{\includegraphics[width=5.6cm,angle=270]{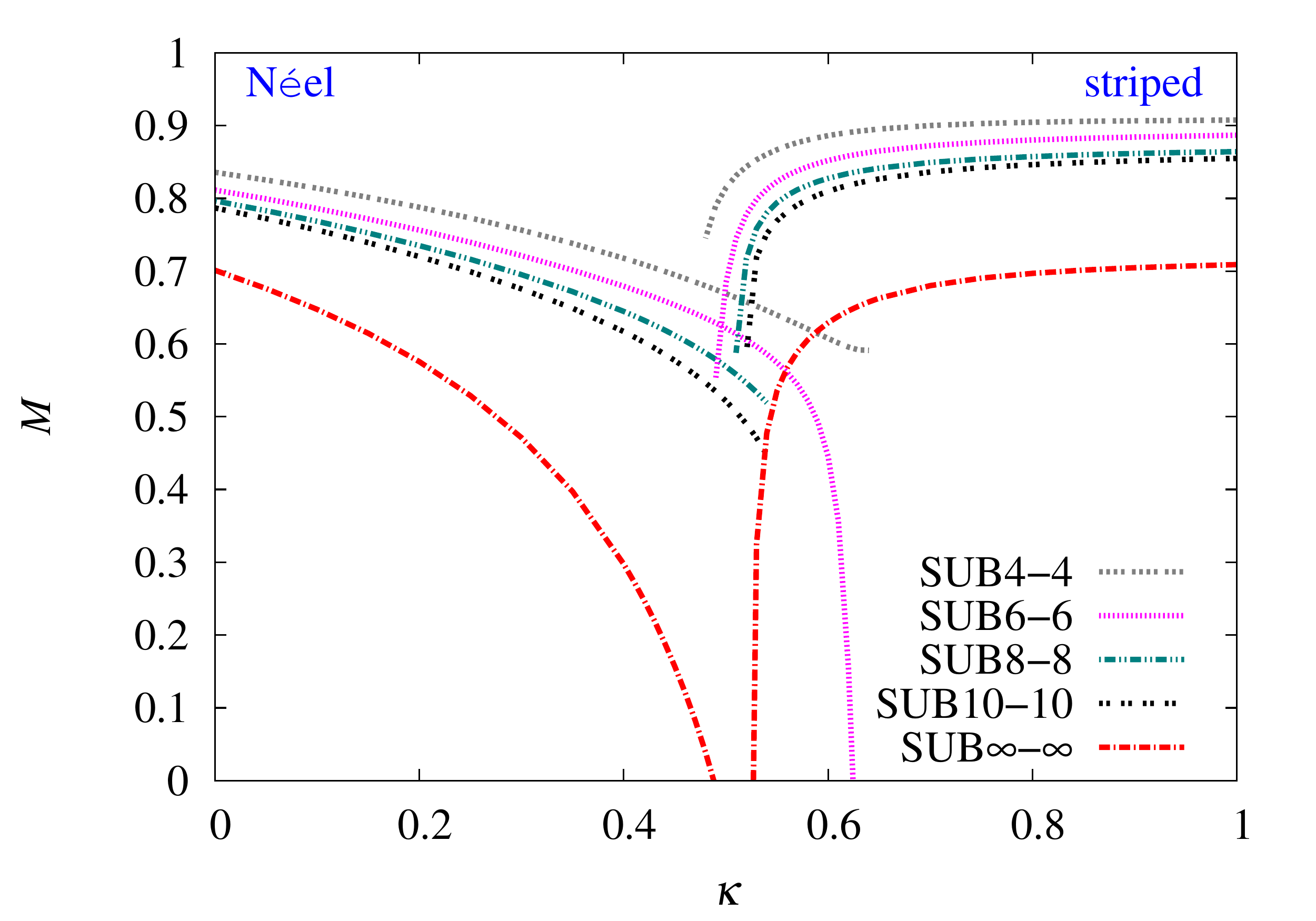}}
\subfigure[]{\includegraphics[width=5.6cm,angle=270]{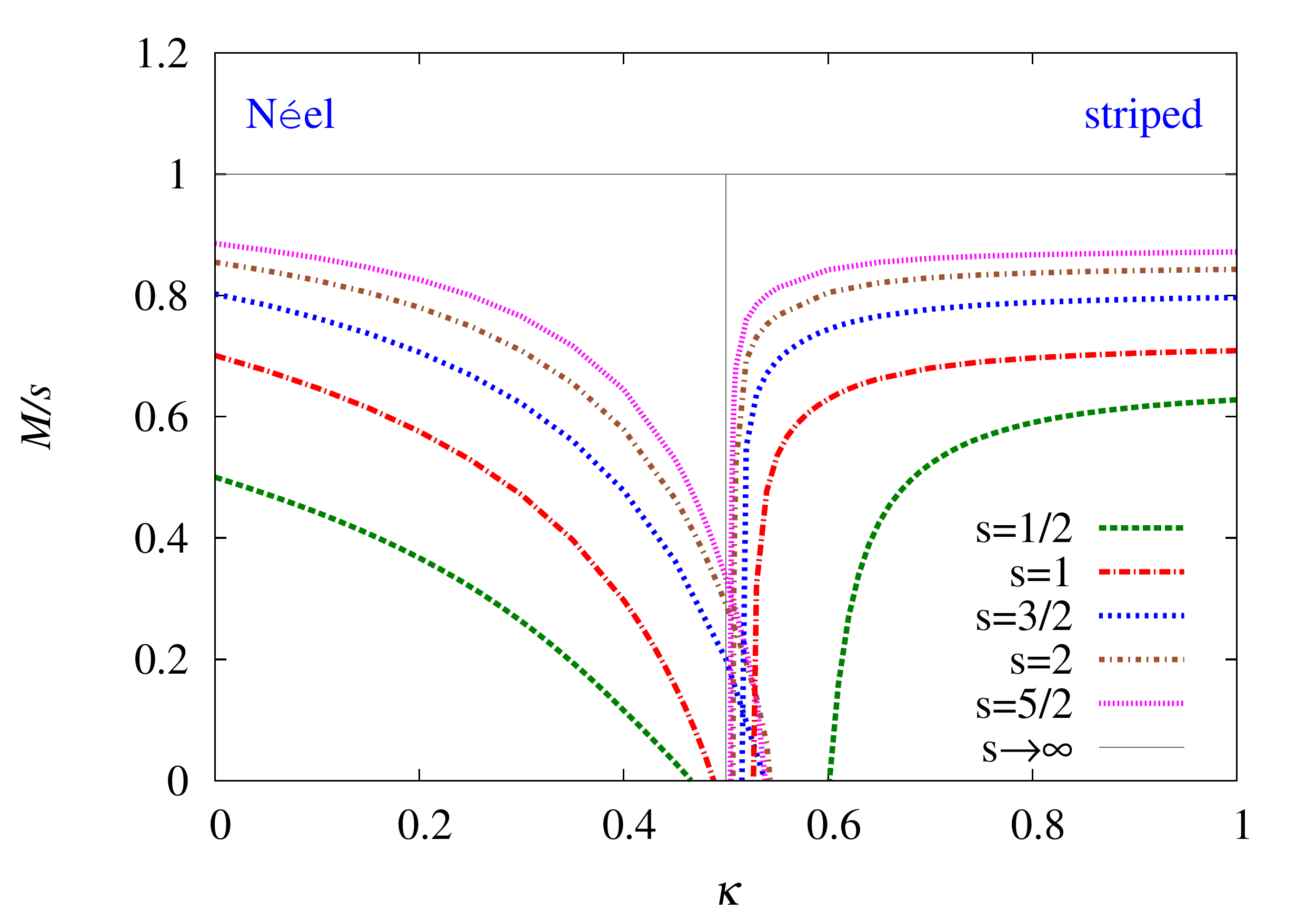}}
  }
\caption{CCM results for the GS magnetic order parameter, $M$, for
  the $J_{1}$--$J_{2}$--$J_{3}$ model on the honeycomb lattice (with $J_{1} \equiv 1, J_{3}=J_{2} \equiv \kappa J_{1} > 0$), as a function of $\kappa$, 
  using the N\'{e}el and striped states as the CCM model states.  (a) The results for the $s=1$ model are shown using the SUB$n$--$n$
  approximations with $n=\{4,6,8,10\}$, together with the corresponding
  extrapolated SUB$\infty$--$\infty$ results obtained using equation (\ref{M_extrapo_frustrated}), with this data set.  (b) We show extrapolated (SUB$\infty$--$\infty$) results for $M/s$ as a function of $\kappa$, using equation
(\ref{M_extrapo_frustrated}) together with the data set $n=\{6,8,10,12\}$ for the case $s=\frac{1}{2}$, and the corresponding data sets with $n=\{4,6,8,10\}$ for the four cases $s=1,\,\frac{3}{2},\,2\,,\frac{5}{2}$.  We also show the corresponding classical ($s \to \infty$) result.}
\label{M_multiSpins}
\end{center}
\end{figure}

In order to get more detailed evidence on the phase structures of the
model for various values of the spin quantum number $s$ we now turn to
the results for the GS magnetic order parameter, $M$, of Eq.\
(\ref{M_eq}).  Thus, firstly, in Fig.\ \ref{M_multiSpins}(a) we show
our CCM results for $M$ for the spin-1 model at various SUB$n$--$n$
levels of approximation with $n=\{4,6,8,10\}$, using both the N\'{e}el
and striped AFM states as separate choices for the model state.
Hence, what is shown in Fig.\ \ref{M_multiSpins}(a) for $M$ is just
the precise analogue of what is shown in Fig.\ \ref{E_multiSpins}(a)
for $E/N$.  It is clear that the SUB$n$--$n$ sequence of
approximations for $M$ converges more slowly than for $E/N$, just as
expected.  We also show in Fig.\ \ref{M_multiSpins}(a) the
extrapolated results for the spin-1 model, where we have used the data
set shown, $n=\{4,6,8,10\}$, as input to the extrapolation scheme of
Eq.\ (\ref{M_extrapo_frustrated}).  As was explained in Sec.\
\ref{ccm_section}, while the alternative scheme of Eq.\
(\ref{M_extrapo_standard}) is certainly more appropriate when the
frustration parameter $\kappa$ is zero or small, that of Eq.\
(\ref{M_extrapo_frustrated}) is certainly preferable for larger values
of $\kappa$ (e.g., in the striped phase) or when the order parameter
$M$ becomes small (i.e., near any QCPs).

The SUB$\infty$--$\infty$ extrapolation shown in Fig.\
\ref{M_multiSpins}(a) now clearly validates the earlier, more
qualitative, results from the GS energy, namely the existence of a GS
phase intermediate between the quasiclassical N\'{e}el and striped
collinear AFM states, just as in the spin-$\frac{1}{2}$ case.  Once
again, N\'{e}el LRO exists over the range $0 \leq \kappa <
\kappa_{c_{1}}$, while striped LRO exists for $\kappa >
\kappa_{c_{2}}$, where $\kappa_{c_{1}} < \kappa_{{\rm cl}} =
\frac{1}{2}$ and $\kappa_{c_{2}} > \kappa_{{\rm cl}} = \frac{1}{2}$.
The values obtained for the two QCPs from the extrapolations using
Eq.\ (\ref{M_extrapo_frustrated}) with the data set $n=\{4,6,8,10\}$,
taken as the points where $M \rightarrow 0$, as shown in Fig.\
\ref{M_multiSpins}(a), are $\kappa_{c_{1}} \approx 0.486$ and
$\kappa_{c_{2}} \approx 0.527$.  A more detailed analysis of the
errors associated with the fits, and by comparison with comparable
extrapolations using alternative data sets (e.g., $n=\{6,8,10\}$ and
$n=\{4,6,8\}$), yields our best estimates for the spin-1 model QCPs,
$\kappa_{c_{1}}=0.485(5)$ and $\kappa_{c_{2}}=0.528(5)$.  These may be
compared with the corresponding values for the spin-$\frac{1}{2}$
model QCPs \cite{DJJF:2011_honeycomb}, $\kappa_{c_{1}}=0.47$ and
$\kappa_{c_{2}}=0.60$.

In Fig.\ \ref{M_multiSpins}(b) we now compare the corresponding
extrapolated curves for the scaled magnetic order parameter, $M/s$,
using both the N\'{e}el and striped AFM states separately as CCM model
states, for the five cases
$s=\frac{1}{2},\,1,\,\frac{3}{2},\,2\,,\frac{5}{2}$.  In each case
shown the extrapolation has been performed with the ansatz of Eq.\
(\ref{M_extrapo_frustrated}), together with the data set
$n=\{6,8,10,12\}$ for the $s=\frac{1}{2}$ case and the sets
$n=\{4,6,8,10\}$ for each case with $s>\frac{1}{2}$.  The results once
again validate our earlier, more qualitative, findings from the GS
energy results, that the intermediate phase is present only for the
two cases $s=\frac{1}{2},\,1$, with a direct transition from the
N\'{e}el to the striped phase in all cases $s > 1$, just as in the
classical $(s \rightarrow \infty)$ limit.  For all cases $s > 1$ this
direct transition clearly occurs at values very close to the classical
value $\kappa_{{\rm cl}} = \frac{1}{2}$.  The actual crossing points
of the order parameter curves shown in Fig.\ \ref{M_multiSpins}(b)
occur at values $\kappa_{m} \approx 0.517$ for $s=\frac{3}{2}$,
$\kappa_{m} \approx 0.508$ for $s=2$, and $\kappa_{m} \approx 0.505$
for $s=\frac{5}{2}$.  What is apparent from Fig.\ \ref{M_multiSpins}(b)
is that the curves for the striped phase approach zero much more steeply
than for the N\'{e}el phase for all values of $s$.  For the case
$s=\frac{1}{2}$ it was argued \cite{DJJF:2011_honeycomb} that this was
a reflection of the transition at $\kappa_{c_{2}}$ being of
first-order type, while that at $\kappa_{c_{1}}$ is of continuous (and
hence probably of the deconfined) type.  The difference in the shapes
of the curves near the crossing point is what leads to the direct
transition apparently being at values slightly larger than
$\kappa_{{\rm cl}} = \frac{1}{2}$ for finite values of $s > 1$.

Clearly the precise crossing points $\kappa_{m}$ of the magnetic order
curves for the N\'{e}el and striped phases for the cases $s>1$ depend
rather critically on the extrapolations, particularly those for the
striped phase, where the slope become large.  In such cases more
precise values of the corresponding QCP for the direct transition
between the two quasiclassical phases can be expected to come from the
analogous crossing points, $\kappa_{e}$, of the extrapolated energy
curves.  The respective values from Fig.\ \ref{E_multiSpins}(b) are
$\kappa_{e} \approx 0.544$ for $s=\frac{3}{2}$, $\kappa_{e} \approx
0.534$ for $s=2$, and $\kappa_{e} \approx 0.528$ for $s=\frac{5}{2}$.
It is reassuring that the respective pairs of values of $\kappa_{e}$
and $\kappa_{m}$ agree so well in each case, for what are essentially
quite independent results.

Before discussing how we can investigate the nature of the
intermediate phase for the present $s=1$ case within the CCM
framework, let us briefly comment on the case of the pure HAF on the
honeycomb lattice, with NN interactions only (i.e., the limiting
case $\kappa=0$ of the present model).  In this case, the
extrapolation ansatz of Eq.\ (\ref{M_extrapo_standard}) becomes
applicable, rather than that of Eq.\ (\ref{M_extrapo_frustrated})
shown in Fig.\ \ref{M_multiSpins}.  We show in Table \ref{GS_E_M} the
scaled values for the GS energy per spin and magnetic order
parameters, $E/(Ns^{2})$ and $M/s$, respectively, for our present
model calculations at the unfrustrated limiting value $\kappa=0$.
\begin{table}   
  \caption{GS parameters of the HAF on the honeycomb lattice, with NN interactions only (of strength $J_{1}=1$), for various values of the spin quantum number $s$.}  
\vspace{0.2cm}
\begin{center}    
\begin{tabular}{ccc}   \hline\hline \\ [-1.5ex]   
$s$ & $E/(Ns^{2})$ & $M/s$   \\ [0.5ex] \hline \\ [-1.7ex]
$\frac{1}{2}$ & -2.17866 & 0.5459 \\ [0.5ex]
1 & -1.83061 & 0.7412 \\  [0.5ex]
$\frac{3}{2}$ & -1.71721 & 0.8249 \\ [0.5ex]
2 & -1.66159 & 0.8689 \\ [0.5ex]
$\frac{5}{2}$ & -1.62862 & 0.8955 \\ [0.5ex] \hline \\ [-1.7ex]
$\infty$ & -1.5 & 1 \\ [0.5ex] \hline\hline
\end{tabular}
\end{center}
\label{GS_E_M}
\end{table}
The corresponding extrapolation schemes of Eqs.\ (\ref{extrapo_E}) and
Eq.\ (\ref{M_extrapo_standard}) have been used in Table \ref{GS_E_M},
together with the input data sets with $n=\{6,8,10,12\}$ for
$s=\frac{1}{2}$, and with $n=\{4,6,8,10\}$ for $s>\frac{1}{2}$.

Another way to estimate the accuracy of our extrapolated CCM results
for the higher spin values is to use them to extract, for example, the
coefficients of the expansions of $E/(Ns^{2})$ and $M/s$ in inverse
powers of $s$, and compare them with the results of higher-order
spin-wave theory (SWT).  For example, at second-order, we may fit our
results of Table \ref{GS_E_M} to the forms,
\begin{equation}
\frac{E}{Ns^{2}} = -\frac{3}{2} + \frac{e_{1}}{s} + \frac{e_{2}}{s^{2}}\,,  \label{honey-pure_E-fit_inversePower}
\end{equation} 
and 
\begin{equation}
\frac{M}{s} = 1 + \frac{m_{1}}{s} + \frac{m_{2}}{s^{2}}\,,   \label{honey-pure_M-fit_inversePower}
\end{equation} 
and then compare with the corresponding results of second-order SWT,
i.e., SWT(2).  If we simply take our results from Table \ref{GS_E_M}
for the two highest spin values calculated, viz., $s=2,\,\frac{5}{2}$,
and fit them to Eqs.\ (\ref{honey-pure_E-fit_inversePower}) and
(\ref{honey-pure_M-fit_inversePower}), we obtain values $e_{1} \approx
-0.31503$ and $e_{2} \approx -0.01630$ for the GS energy, and $m_{1}
\approx -0.2575$ and $m_{2} \approx -0.0095$ for the GS N\'{e}el
magnetic order parameter (i.e., the sublattice magnetization).  The
corresponding (exact) SWT(2) results \cite{Zheng:1991_honey} are
$e_{1} = -0.31476$, $e_{2} = -0.01651$, $m_{1} = -0.2582$, and $m_{2}
= 0$.  The agreement is rather striking.

We now turn finally to the question of what is the nature of the
intermediate phase in the case $s=1$.  For the analogous
$s=\frac{1}{2}$ case it was shown \cite{DJJF:2011_honeycomb} that the
intermediate paramagnetic phase likely had PVBC order.  It is natural
now to consider this possibility for the $s=1$ case.  To do so we now
calculate within the CCM framework the susceptibility, $\chi_{p}$,
which measures the response of the system to an applied external field
that promotes PVBC order.  More generally, let us add an infinitesimal
field operator $F \equiv \delta\; \hat{O}$ to the Hamiltonian $H$ of Eq.\
(\ref{Hamiltonian}).  We then calculate the perturbed energy per site,
$E(\delta)/N=e(\delta)$, for the perturbed Hamiltonian $H+F$, using
the same CCM procedure as above, and using the same previous model
states.  The susceptibility of the system to the perturbed operator
$\hat{O}$ is then defined as usual to be
\begin{equation}
\chi_{F} \equiv -
\left. \frac{\partial^2e(\delta)}{\partial {\delta}^2} \right|_{\delta=0\,,}
\end{equation} 
so that the energy,
\begin{equation}
e(\delta) = e(0) - \frac{1}{2}\chi_{F}\delta^{2}\,,
\end{equation}
is a maximum at $\delta=0$ for $\chi_{F}>0$.  A clear signal of the
system becoming unstable against the perturbation $F$ is the finding
that $\chi_{F}$ diverges or, equivalently, that $\chi_{F}^{-1}$
becomes zero (and then possibly changes sign).

\begin{figure}
\begin{center}
\mbox{
\raisebox{-5cm}{  \subfigure[]{\includegraphics[width=2.2cm,height=2.2cm]{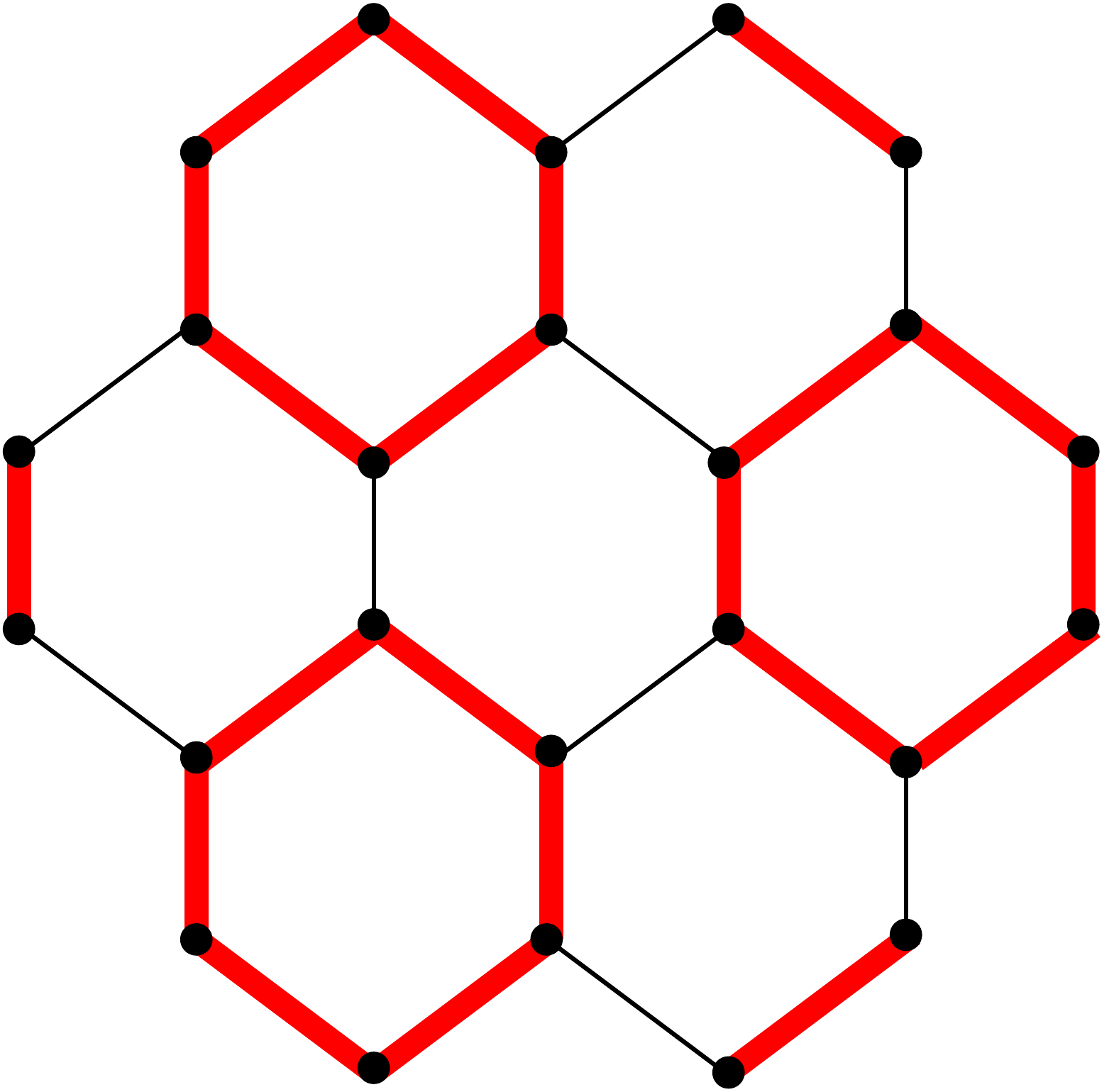}}}
\hspace{-1cm} \subfigure[]{\includegraphics[width=6cm,angle=270]{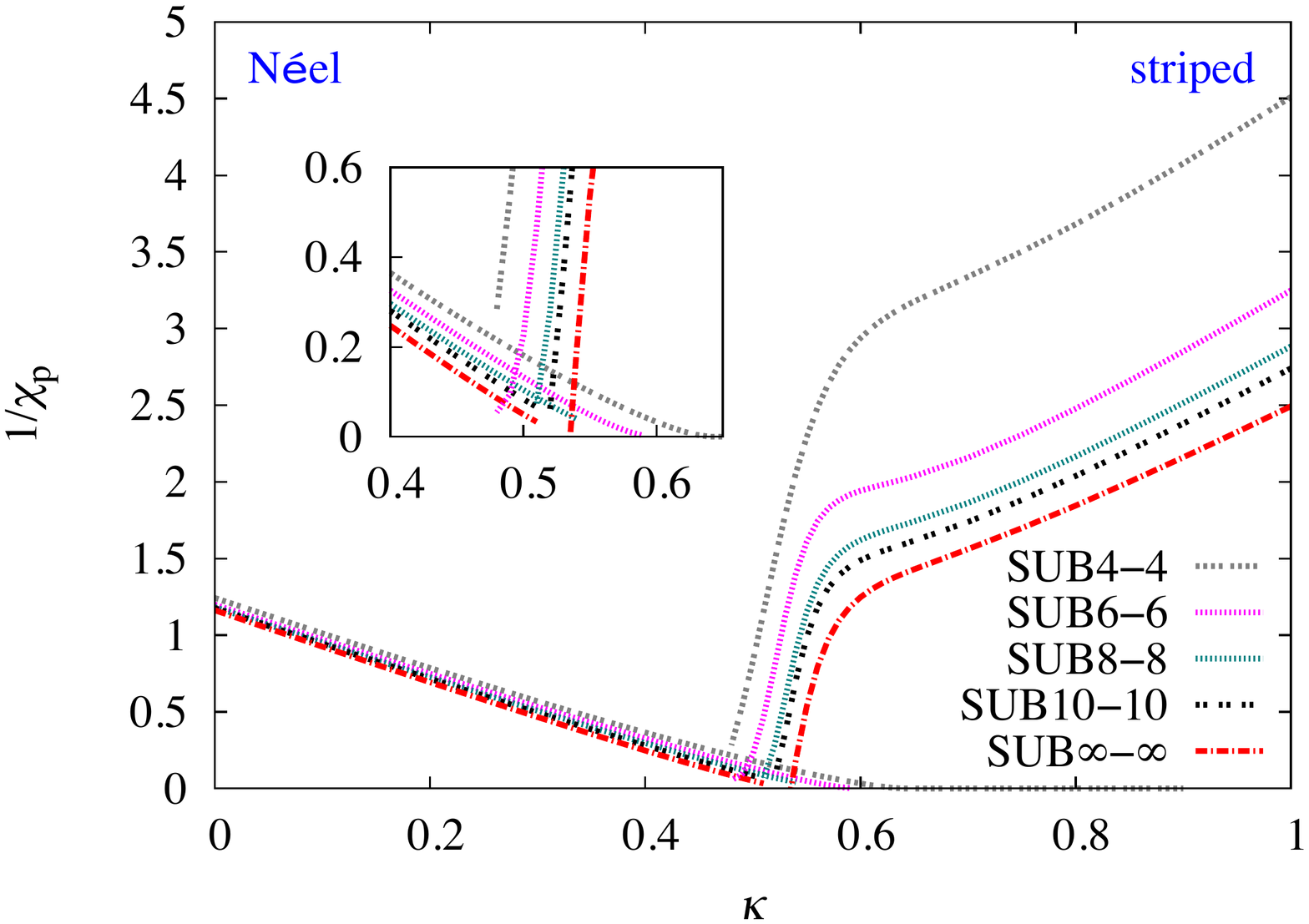}}
\hspace{-2cm} \subfigure[]{\includegraphics[width=6cm,angle=270]{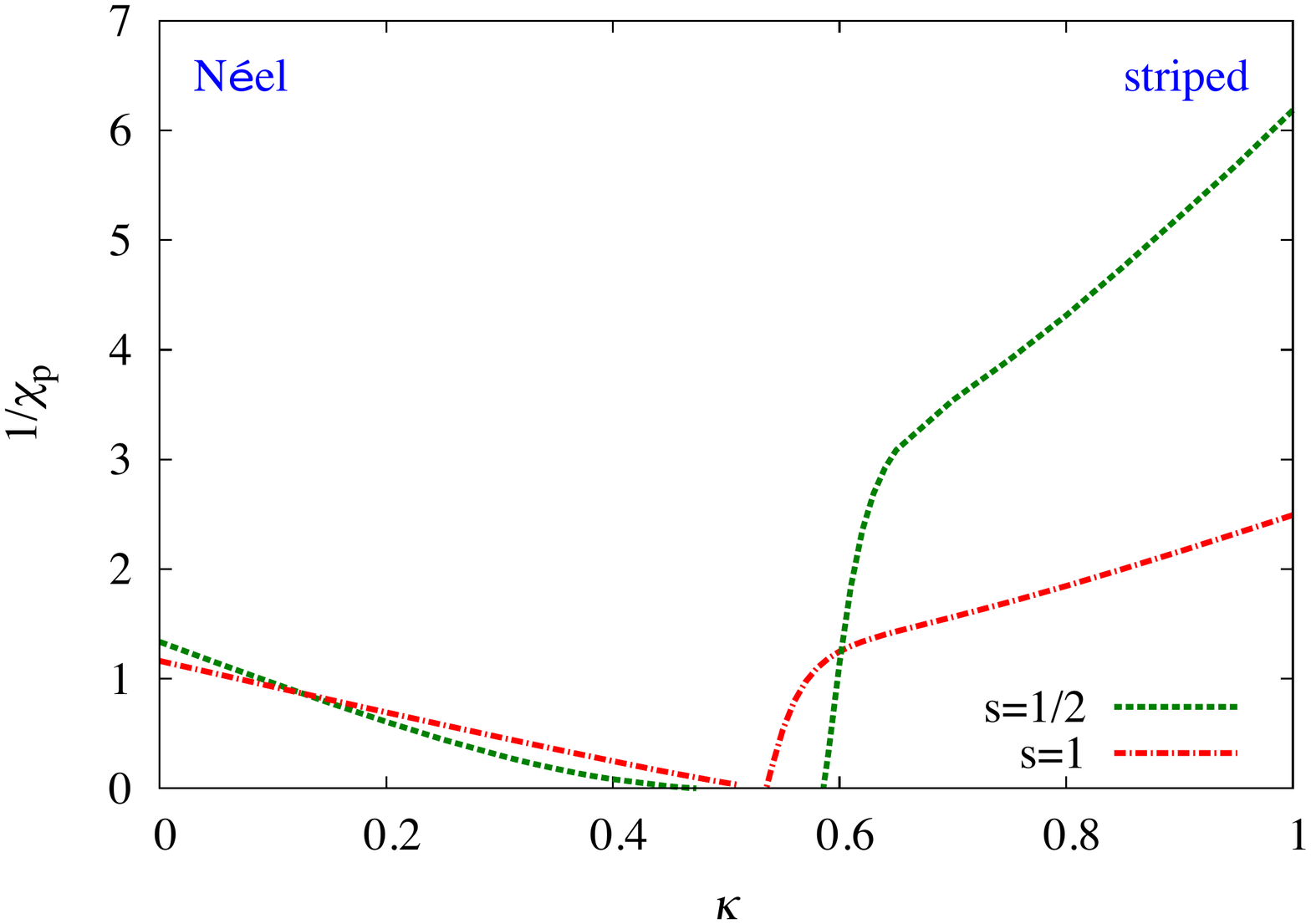}}
  }
\caption{(a) The 
  fields $F=\delta\; \hat{O}$ for the plaquette susceptibility
  $\chi_{p}$.  Thick (red) and thin (black) lines correspond respectively
  to strengthened and weakened NN exchange couplings, where $\hat{O} =
  \sum_{\langle i,j \rangle} a_{ij}
  \mathbf{s}_{i}\cdot\mathbf{s}_{j}$, and the sum runs over all NN
  bonds, with $a_{ij}=+1$ and $-1$ for thick (red) and thin (black)
  lines respectively.  (b) CCM results for the inverse plaquette
  susceptibility, $1/\chi_{p}$, for
  the $J_{1}$--$J_{2}$--$J_{3}$ model on the honeycomb lattice (with $J_{1} \equiv 1, J_{3}=J_{2} \equiv \kappa J_{1} > 0$), as a function of $\kappa$, 
  using the N\'{e}el and striped states as the CCM model states.  The results for the $s=1$ model are shown using the SUB$n$--$n$
  approximations with $n=\{4,6,8,10\}$, together with the corresponding
  extrapolated SUB$\infty$--$\infty$ results obtained using equation (\ref{X_extrapo}), with this data set.  (c) We show extrapolated (SUB$\infty$--$\infty$) results for $1/\chi_{p}$ as a function of $\kappa$, using equation
(\ref{X_extrapo}) together with the data set $n=\{6,8,10,12\}$ for the case $s=\frac{1}{2}$, and the data set $n=\{4,6,8,10\}$ for the case $s=1$.
}
\label{X_multiSpins}
\end{center}
\end{figure}

In our present case the perturbing operator $F$ is now chosen to
promote PVBC order, and it is illustrated in Fig.\
\ref{X_multiSpins}(a).  
It clearly breaks the translational symmetry
of the system.  The SUB$n$--$n$ estimates, $\chi_{p}(n)$, for the
resulting susceptibility of our system to the formation of PVBC order,
then need to be extrapolated to the exact ($n \rightarrow \infty$)
limit.  Previous experience \cite{DJJF:2011_honeycomb} has shown that
an appropriate extrapolation ansatz is
\begin{equation}
\chi_{p}^{-1}(n) = x_{0}+x_{1}n^{-2}+x_{2}n^{-4}\,.  \label{X_extrapo}
\end{equation}

We show in Fig.\ \ref{X_multiSpins}(b) our CCM results for
$\chi_{p}^{-1}(n)$ at SUB$n$--$n$ levels of approximation with
$n=\{4,6,8,10\}$, using both the N\'{e}el and striped AFM states
separately as model states, in complete analogy to what is displayed
in Figs.\ \ref{E_multiSpins}(a) and \ref{M_multiSpins}(a) for the GS
energy per spin, $E/N$, and magnetic order parameter, $M$.  In Fig.\
\ref{X_multiSpins}(b) we also show the extrapolated results ($x_{0}$)
obtained by inserting the set of raw results shown into Eq.\
(\ref{X_extrapo}).  Just as in the spin-$\frac{1}{2}$ case
\cite{DJJF:2011_honeycomb}, the results for $\chi_{p}^{-1}(n)$
converge much faster for the N\'{e}el state than for the striped
state.  For both states there are clear critical points at which
$\chi_{p}^{-1}$ vanishes.  However, the shapes of the curves for
$\chi_{p}^{-1}$ near their respective critical points differ markedly,
just as in the spin-$\frac{1}{2}$ case.  Thus, on the N\'{e}el side,
$\chi_{p}^{-1} \rightarrow 0$ with a slope that is small.  By
contrast, on the striped side, $\chi_{p}^{-1} \rightarrow 0$ with a
very large slope (and which is probably compatible with being
infinite, within extrapolation errors).  These differences reinforce
our earlier findings that the critical point at which N\'{e}el order
vanishes is likely to mark a continuous phase transition, while that
at which striped order vanishes is likely to mark a first-order
transition.

In Fig.\ \ref{X_multiSpins}(c) we compare the extrapolated CCM results
for $\chi_{p}^{-1}$ for the two cases ($s=\frac{1}{2},\,1$) for which
our findings indicate the existence in the model of a phase in its
$T=0$ GS phase diagram intermediate between the two quasiclassical
phases with magnetic LRO.  Whereas in the spin-$\frac{1}{2}$ case
there exists a clear gap along the frustration parameter, $\kappa$,
axis between the two points at which $\chi_{p}^{-1}$ vanishes (one for
each quasiclassical phase), the gap in the spin-1 case is much less
marked.  Indeed, its very existence is open to doubt, as we explain
below.  For the spin-$\frac{1}{2}$ case \cite{DJJF:2011_honeycomb} the
two values on Fig.\ \ref{X_multiSpins}(c) at which $\chi_{p}^{-1}
\rightarrow 0$ are $\kappa \approx 0.473$ and $\kappa \approx 0.586$.
These may be compared with the corresponding values on Fig.\
\ref{M_multiSpins}(b) at which $M \rightarrow 0$, which are
$\kappa_{c_{1}} \approx 0.466$ and $\kappa_{c_{2}} \approx 0.601$.
The very close agreement between the corresponding values was taken
\cite{DJJF:2011_honeycomb} to be good evidence that the PVBC phase
occurs at (or is very close to) the transition points $\kappa_{c_{1}}$
and $\kappa_{c_{2}}$ where the quasiclassical magnetic LRO vanishes.
The fact that the slope of the $\chi_{p}^{-1}(\kappa)$ curve on the
N\'{e}el side is vanishingly small (within numerical errors) at the
point where $\chi_{p}^{-1} \rightarrow 0$, also provided strong
evidence that $\chi_{p}^{-1}$ vanishes over the entire intermediate
region in Fig.\ \ref{X_multiSpins}(c) between the points, where the
CCM calculations have been performed with the two classical model
states.  All of this evidence pointed strongly to the stable GS phase
in the whole of the region $\kappa_{c_{1}} < \kappa < \kappa_{c_{2}}$
being one with PVBC order, in the $s=\frac{1}{2}$ case.  If any other
phase exists in part of this region, its region of stability is
clearly constrained by the CCM calculations \cite{DJJF:2011_honeycomb}
to be a very small part of the intermediate region.

The corresponding situation for the $s=1$ case is now subtly
different, however.  Thus, firstly, a close inspection of Fig.\
\ref{X_multiSpins}(b) (and, especially, the inset) shows that on the
N\'{e}el side the extrapolated SUB$\infty$--$\infty$
$\chi_{p}^{-1}(\kappa)$ curve has not quite reached zero at the
SUB10--10 termination point (or at least as far as we have managed to
perform numerical calculations, which, as we discussed above, become
increasingly difficult and computationally costly the closer one
approaches a termination point).  A simple further extrapolation of
the curve, however, yields a value $\kappa \approx 0.530(2)$ at which
$\chi_{p}^{-1} \rightarrow 0$ on the N\'{e}el side.  This is very close to the corresponding value in Fig.\
\ref{X_multiSpins}(b) of $\kappa \approx 0.535$ at which
$\chi_{p}^{-1} \rightarrow 0$ on the striped phase side.
Significantly, both of these values are indubitably greater than
$\kappa_{{\rm cl}} = \frac{1}{2}$.  By contrast, the corresponding
values from Fig.\ \ref{M_multiSpins}(b) at which $M \rightarrow 0$ for
the spin-1 model are $\kappa_{c_{1}} \approx 0.486$ and
$\kappa_{c_{2}} \approx 0.527$.  The most likely interpretation of our
$\chi_{p}$ results is hence that the CCM using both model states is
showing that $\chi_{p}^{-1}$ vanishes only at a single point, viz.,
the QCP $\kappa_{c_{2}}$, for the spin-1 model.  This interpretation
is lent further weight by the observation that, unlike in the
spin-$\frac{1}{2}$ case, the slope of the $\chi_{p}^{-1}(\kappa)$
curve on the N\'{e}el side does not appear to be zero at the
(extrapolated) point at which it becomes zero.

In summary, if a PVBC-ordered phase is stable anywhere in the
intermediate regime $\kappa_{c_{1}} < \kappa < \kappa_{c_{2}}$, our
findings are that it is confined only to a very narrow range close to
$\kappa_{c_{2}}$, and that it is definitely not the stable GS phase
over the whole interval.  The more likely scenario is that the
intermediate regime is occupied by a paramagnetic phase (or more than
one such phase) with a form of order other than PVBC.

\section{Conclusions}
\label{conclusions_sec}
We have investigated higher-spin versions of a frustrated
$J_{1}$--$J_{2}$--$J_{3}$ HAF model on the honeycomb lattice, in the
specific case where $J_{1} > 0$, $J_{3}=J_{2} \equiv \kappa J_{1}>0$,
over the range $0 \leq \kappa \leq 1$ of the frustration parameter.
This includes the point of maximum frustration in the classical ($s
\rightarrow \infty$) limit, viz., the tricritical point at
$\kappa_{{\rm cl}} = \frac{1}{2}$, at which there is a direct
first-order transition (along the line $J_{3}=J_{2}$) between a N\'{e}el-ordered AFM phase for
$\kappa < \kappa_{{\rm cl}}$ and a collinear stripe-ordered AFM phase
for $\kappa > \kappa_{{\rm cl}}$.  Whereas the spin-$\frac{1}{2}$
version of the model has been studied previously, higher-spin versions
have received no attention, so far as we are aware.

In particular, the CCM has been applied to the spin-$\frac{1}{2}$
model in an earlier study \cite{DJJF:2011_honeycomb} that yielded
accurate results for its entire $T=0$ GS phase diagram.  Since the method
generally provides values for the QCPs of a wide range of spin-lattice
systems, which are among the most accurate available by any
alternative methodology, we have now used the CCM to study spin-$s$
versions of the model with values of the spin quantum number
$s>\frac{1}{2}$.  An aim has been to compare the $T=0$ GS phase
diagrams of the higher-spin models with the two extreme limits,
$s=\frac{1}{2}$ (where quantum effects should be greatest) and $s
\rightarrow \infty$ (where quantum effects vanish).  In particular, it
has been shown \cite{DJJF:2011_honeycomb} that the direct classical
transition at $\kappa_{{\rm cl}} = \frac{1}{2}$ is split by quantum
fluctuations in the spin-$\frac{1}{2}$ model into two separate
transitions at $\kappa_{c_{1}} < \kappa_{{\rm cl}}$ and
$\kappa_{c_{2}} > \kappa_{{\rm cl}}$, at the first of which N\'{e}el
AFM order of the type shown in Fig.\ \ref{model_pattern}(a) breaks
down, and at the second of which striped AFM order of the type shown
in Fig.\ \ref{model_pattern}(b) breaks down.  Between the regimes
$\kappa < \kappa_{c_{1}}$ in which the stable GS phase has N\'{e}el
magnetic LRO and $\kappa > \kappa_{c_{2}}$ in which the stable GS
phase has striped magnetic LRO, there opens an intermediate
paramagnetic regime $\kappa_{c_{1}} < \kappa < \kappa_{c_{2}}$.
Strong evidence was presented \cite{DJJF:2011_honeycomb} that the
stable GS phase in this entire intermediate regime in the
spin-$\frac{1}{2}$ case is an intrinsically quantum-mechanical one
with PVBC order.

We have now performed analogous CCM calculations for spin-$s$ versions
of the same honeycomb lattice model for values
$s=1,\,\frac{3}{2},\,2\,,\frac{5}{2}$.  A primary finding is that an
intermediate phase also exists for the case $s=1$, but that for all
higher spins ($s \geq \frac{3}{2}$) the intermediate phase disappears
in favour of a direct transition between the two quasiclassical states
with magnetic LRO.  For all finite values of the spin quantum number
$s \geq \frac{3}{2}$ the direct transition seems to occur at a value
$\kappa_{c}$ marginally higher than the classical value of 0.5 [e.g.,
for $s=\frac{3}{2}$, $\kappa_{c} \approx 0.53(1)$, and for
$s=\frac{5}{2}$, $\kappa_{c} \approx 0.52(1)$], with $\kappa_{c}$
tending monotonically to 0.5 as $s \rightarrow \infty$.

The range of the intermediate phase is smaller for the spin-1 model
($\kappa_{c_{1}} \approx 0.49$,\,$\kappa_{c_{2}} \approx 0.53$) than
for its spin-$\frac{1}{2}$ counterpart ($\kappa_{c_{1}} \approx
0.47$,\,$\kappa_{c_{2}} \approx 0.60$), as expected.  Interestingly,
all of the evidence garnered here is that, unlike in the
spin-$\frac{1}{2}$ case the intermediate phase for the spin-1 model
does {\it not} have PVBC ordering.  On the other hand, both the
spin-$\frac{1}{2}$ and spin-1 models seem to share that the transition
at $\kappa_{c_{2}}$ is a direct first-order one while that at
$\kappa_{c_{1}}$ is continuous.

On the basis of the present calculations, the nature and properties of
the intermediate phase in the spin-1 version of the model remain
open questions.  It will be of considerable interest to study this phase
further, both by the CCM and by the use of alternative techniques.  A
particularly promising such alternative technique in this respect is
the DMRG method, which has been used very recently
\cite{Gong:2015_honey_J1J2mod_s1} in an analysis of the quantum
($T=0$) phase diagram of the spin-1 version of the related
$J_{1}$--$J_{2}$ Heisenberg model on the honeycomb lattice.

\section*{Acknowledgments} 
We thank the University of Minnesota Supercomputing Institute for the
grant of supercomputing facilities.  One of us (RFB) also gratefully
acknowledges the Leverhulme Trust for the award of an Emeritus
Fellowship (EM-2015-07).

\section*{References} 

\bibliographystyle{iopart-num}
\bibliography{bib_general}

\providecommand{\noopsort}[1]{}\providecommand{\singleletter}[1]{#1}%
\providecommand{\newblock}{}
\begin{thebibliography}{10}
\expandafter\ifx\csname url\endcsname\relax
  \def\url#1{{\tt #1}}\fi
\expandafter\ifx\csname urlprefix\endcsname\relax\def\urlprefix{URL }\fi
\providecommand{\eprint}[2][]{\url{#2}}

\bibitem{Mermin:1966}
Mermin N~D and Wagner H 1966 {\em Phys. Rev. Lett.\/} {\bf 17} 1133

\bibitem{Rastelli:1979_honey}
Rastelli E, Tassi A and Reatto L 1979 {\em Physica B \& C\/} {\bf 97} 1

\bibitem{Fouet:2001_honey}
Fouet J~B, Sindzingre P and Lhuillier C 2001 {\em Eur. Phys. J. B\/} {\bf 20}
  241

\bibitem{Cabra:2011_honey}
Cabra D~C, Lamas C~A and Rosales H~D 2011 {\em Phys. Rev. B\/} {\bf 83} 094506

\bibitem{Mattsson:1994_honey}
Mattsson A, Fr{\"{o}}jdh P and Einarsson T 1994 {\em Phys. Rev. B\/} {\bf 49}
  3997--4002

\bibitem{Mulder:2010_honey}
Mulder A, Ganesh R, Capriotti L and Paramekanti A 2010 {\em Phys. Rev. B\/}
  {\bf 81} 214419

\bibitem{Ganesh:2011_honey}
Ganesh R, Sheng D~N, Kim Y~J and Paramekanti A 2011 {\em Phys. Rev. B\/} {\bf
  83} 144414

\bibitem{Clark:2011_honey}
Clark B~K, Abanin D~A and Sondhi S~L 2011 {\em Phys. Rev. Lett.\/} {\bf 107}
  087204

\bibitem{DJJF:2011_honeycomb}
Farnell D~J~J, Bishop R~F, Li P~H~Y, Richter J and Campbell C~E 2011 {\em Phys.
  Rev. B\/} {\bf 84} 012403

\bibitem{Reuther:2011_honey}
Reuther J, Abanin D~A and Thomale R 2011 {\em Phys. Rev. B\/} {\bf 84} 014417

\bibitem{Albuquerque:2011_honey}
Albuquerque A~F, Schwandt D, Het{\'e}nyi B, Capponi S, Mambrini M and
  L{\"{a}}uchli A~M 2011 {\em Phys. Rev. B\/} {\bf 84} 024406

\bibitem{Mosadeq:2011_honey}
Mosadeq H, Shahbazi F and Jafari S~A 2011 {\em J. Phys.: Condens. Matter\/}
  {\bf 23} 226006

\bibitem{Oitmaa:2011_honey}
Oitmaa J and Singh R~R~P 2011 {\em Phys. Rev. B\/} {\bf 84} 094424

\bibitem{Mezzacapo:2012_honey}
Mezzacapo F and Boninsegni M 2012 {\em Phys. Rev. B\/} {\bf 85} 060402(R)

\bibitem{PHYLi:2012_honeycomb_J1neg}
Li P~H~Y, Bishop R~F, Farnell D~J~J, Richter J and Campbell C~E 2012 {\em Phys.
  Rev. B\/} {\bf 85} 085115

\bibitem{Bishop:2012_honeyJ1-J2}
Bishop R~F, Li P~H~Y, Farnell D~J~J and Campbell C~E 2012 {\em J. Phys.:
  Condens. Matter\/} {\bf 24} 236002

\bibitem{Bishop:2012_honey_phase}
Bishop R~F and Li P~H~Y 2012 {\em Phys. Rev. B\/} {\bf 85} 155135

\bibitem{Li:2012_honey_full}
Li P~H~Y, Bishop R~F, Farnell D~J~J and Campbell C~E 2012 {\em Phys. Rev. B\/}
  {\bf 86} 144404

\bibitem{RFB:2013_hcomb_SDVBC}
Bishop R~F, Li P~H~Y and Campbell C~E 2013 {\em J. Phys.: Condens. Matter\/}
  {\bf 25} 306002

\bibitem{Ganesh:2013_honey_J1J2mod-XXX}
Ganesh R, van~den Brink J and Nishimoto S 2013 {\em Phys. Rev. Lett.\/} {\bf
  110} 127203

\bibitem{Zhu:2013_honey_XY}
Zhu Z, Huse D~A and White S~R 2013 {\em Phys. Rev. Lett.\/} {\bf 111} 257201

\bibitem{Zhang:2013_honey}
Zhang H and Lamas C~A 2013 {\em Phys. Rev. B\/} {\bf 87} 024415

\bibitem{Gong:2013_J1J2mod-XXX}
Gong S~S, Sheng D~N, Motrunich O~I and Fisher M~P~A 2013 {\em Phys. Rev. B\/}
  {\bf 88} 165138

\bibitem{Yu:2014_honey_J1J2mod}
Yu X~L, Liu D~Y, Li P and Zou L~J 2014 {\em Physica E\/} {\bf 59} 41

\bibitem{Gong:2015_honey_J1J2mod_s1}
Gong S~S, Zhu W and Sheng D~N 2015 Quantum phase diagram of the spin-1
  ${J}_{1}$--${J}_{2}$ {H}eisenberg model on the honeycomb lattice
  arXiv:1508.00515

\bibitem{Miura:2006_honey}
Miura Y, Hirai R, Kobayashi Y and Sato M 2006 {\em J. Phys. Soc. Jpn.\/} {\bf
  75} 084707

\bibitem{Kataev:2005_honey}
Kataev V, M{\"{o}}ller A, L{\"{o}}w U, Jung W, Schittner N, Kriener M and
  Freimuth A 2005 {\em J. Magn. Magn. Mater.\/} {\bf 290--291} 310--313

\bibitem{Tsirlin:2010_honey}
Tsirlin A~A, Janson O and Rosner H 2010 {\em Phys. Rev. B\/} {\bf 82} 144416

\bibitem{Climent:2012_honey}
Climent-Pascual E, Norby P, Andersen N, Stephens P, Zandbergen H, Larsen J and
  Cava R 2012 {\em Inorg. Chem.\/} {\bf 51} 557--565

\bibitem{Singh:2010_honey}
Singh Y and Gegenwart P 2010 {\em Phys. Rev. B\/} {\bf 82} 064412

\bibitem{Liu:2011_honey}
Liu X, Berlijn T, Yin W~G, Ku W, Tsvelik A, Kim Y~J, Gretarsson H, Singh Y,
  Gegenwart P and Hill J~P 2011 {\em Phys. Rev. B\/} {\bf 83} 220403(R)

\bibitem{Singh:2012_honey}
Singh Y, Manni S, Reuther J, Berlijn T, Thomale R, Ku W, Trebst S and Gegenwart
  P 2012 {\em Phys. Rev. Lett.\/} {\bf 108} 127203

\bibitem{Choi:2012_honey}
Choi S~K, Coldea R, Kolmogorov A~N, Lancaster T, Mazin I~I, Blundell S~J,
  Radaelli P~G, Singh Y, Gegenwart P, Choi K~R, Cheong S~W, Baker P~J, Stock C
  and Taylor J 2012 {\em Phys. Rev. Lett.\/} {\bf 108} 127204

\bibitem{Regnault:1990_honey}
Regnault L~P and Rossat-Mignod J 1990 {\em Magnetic properties of layered
  transition metal compounds\/} ed {De Jongh} L~J (Dordrecht: Kluwer Academic
  Publishers) pp 271--321

\bibitem{Roudebush:2013_honey}
Roudebush J~H, Andersen N~H, Ramlau R, Garlea V~O, Toft-Petersen R, Norby P,
  Schneider R, Hay J~N and Cava R~J 2013 {\em Inorg. Chem.\/} {\bf 52}
  6083--6095

\bibitem{Smirnova:2009:honey_spin_3half}
Smirnova O, Azuma M, Kumada N, Kusano Y, Matsuda M, Shimakawa Y, Takei T,
  Yonesaki Y and Kinomura N 2009 {\em J. Am. Chem. Soc.\/} {\bf 131} 8313--8317

\bibitem{Okubo:2010:honey_spin_3half}
Okubo S, Elmasry F, Zhang W, Fujisawa M, Sakurai T, Ohta H, Azuma M, Sumirnova
  O~A and Kumada N 2010 {\em J. Phys.: Conf. Ser.\/} {\bf 200} 022042

\bibitem{Villain:1977_ordByDisord}
Villain J 1977 {\em J. Phys. (France)\/} {\bf 38} 385

\bibitem{Villain:1980_ordByDisord}
Villain J, Bidaux R, Carton J~P and Conte R 1980 {\em J. Phys. (France)\/} {\bf
  41} 1263

\bibitem{Shender:1982_ordByDisord}
Shender E~F 1982 {\em Zh. Eksp. Teor. Fiz.\/} {\bf 83} 326--337

\bibitem{Senthil:2004_Science_deconfinedQC}
Senthil T, Vishwanath A, Balents L, Sachdev S and Fisher M~P~A 2004 {\em
  Science\/} {\bf 303} 1490

\bibitem{Senthil:2004_PRB_deconfinedQC}
Senthil T, Balents L, Sachdev S, Vishwanath A and Fisher M~P~A 2004 {\em Phys.
  Rev. B\/} {\bf 70} 144407

\bibitem{Bishop:1978_ccm}
Bishop R~F and L{\"{u}}hrmann K~H 1978 {\em Phys. Rev. B\/} {\bf 17} 3757--3780

\bibitem{Bishop:1982_ccm}
Bishop R~F and L{\"{u}}hrmann K~H 1982 {\em Phys. Rev. B\/} {\bf 26} 5523--5557

\bibitem{Arponen:1983_ccm}
Arponen J 1983 {\em Ann. Phys. (N.Y.)\/} {\bf 151} 311--382

\bibitem{Bishop:1987_ccm}
Bishop R~F and K{\"{u}}mmel H~G 1987 {\em Phys. Today\/} {\bf 40(3)} 52

\bibitem{Arponen:1991_ccm}
Arponen J~S and Bishop R~F 1991 {\em Ann. Phys. (N.Y.)\/} {\bf 207} 171

\bibitem{Bishop:1991_TheorChimActa_QMBT}
Bishop R~F 1991 {\em Theor. Chim. Acta\/} {\bf 80} 95

\bibitem{Bishop:1998_QMBT_coll}
Bishop R~F 1998 {\em Microscopic Quantum Many-Body Theories and Their
  Applications\/} Lecture Notes in Physics Vol. 510 ed Navarro J and Polls A
  (Berlin: Springer-Verlag) p~1

\bibitem{Fa:2004_QM-coll}
Farnell D~J~J and Bishop R~F 2004 {\em Quantum Magnetism\/} Lecture Notes in
  Physics Vol. 645 ed Schollw{\"{o}}ck U, Richter J, Farnell D~J~J and Bishop
  R~F (Berlin: Springer-Verlag) p 307

\bibitem{Bartlett:1989_ccm}
Bartlett R~J 1989 {\em J. Phys. Chem.\/} {\bf 93} 1697

\bibitem{Kummel:1978_ccm}
K{\"{u}}mmel H, L{\"{u}}hrmann K~H and Zabolitzky J~G 1978 {\em Phys Rep.\/}
  {\bf 36C} 1

\bibitem{Zeng:1998_SqLatt_TrianLatt}
Zeng C, Farnell D~J~J and Bishop R~F 1998 {\em J. Stat. Phys.\/} {\bf 90} 327

\bibitem{ccm_code}
We use the program package CCCM of D.~J.~J. Farnell and J.~Schulenburg, see
  http://www-e.uni-magdeburg.de/jschulen/ccm/index.html

\bibitem{Bishop:2000_XXZ}
Bishop R~F, Farnell D~J~J, Kr{\"{u}}ger S~E, Parkinson J~B, Richter J and Zeng
  C 2000 {\em J. Phys.: Condens. Matter\/} {\bf 12} 6887

\bibitem{Kruger:2000_JJprime}
Kr{\"{u}}ger S~E, Richter J, Schulenburg J, Farnell D~J~J and Bishop R~F 2000
  {\em Phys. Rev. B\/} {\bf 61} 14607

\bibitem{Fa:2001_SqLatt_s1}
Farnell D~J~J, Gernoth K~A and Bishop R~F 2001 {\em Phys. Rev. B\/} {\bf 64}
  172409

\bibitem{Darradi:2005_Shastry-Sutherland}
Darradi R, Richter J and Farnell D~J~J 2005 {\em Phys. Rev. B\/} {\bf 72}
  104425

\bibitem{Darradi:2008_J1J2mod}
Darradi R, Derzhko O, Zinke R, Schulenburg J, Kr{\"{u}}ger S~E and Richter J
  2008 {\em Phys. Rev. B\/} {\bf 78} 214415

\bibitem{Bi:2008_EPL_J1J1primeJ2_s1}
Bishop R~F, Li P~H~Y, Darradi R and Richter J 2008 {\em Europhys. Lett.\/} {\bf
  83} 47004

\bibitem{Bi:2008_JPCM_J1xxzJ2xxz_s1}
Bishop R~F, Li P~H~Y, Darradi R, Richter J and Campbell C~E 2008 {\em J. Phys.:
  Condens. Matter\/} {\bf 20} 415213

\bibitem{Bi:2009_SqTriangle}
Bishop R~F, Li P~H~Y, Farnell D~J~J and Campbell C~E 2009 {\em Phys. Rev. B\/}
  {\bf 79} 174405

\bibitem{Bishop:2010_UJack}
Bishop R~F, Li P~H~Y, Farnell D~J~J and Campbell C~E 2010 {\em Phys. Rev. B\/}
  {\bf 82} 024416

\bibitem{Bishop:2010_KagomeSq}
Bishop R~F, Li P~H~Y, Farnell D~J~J and Campbell C~E 2010 {\em Phys. Rev. B\/}
  {\bf 82} 104406

\bibitem{Bishop:2011_UJack_GrtSpins}
Bishop R~F and Li P~H~Y 2011 {\em Eur. Phys. J. B\/} {\bf 81} 37

\bibitem{PHYLi:2012_SqTriangle_grtSpins}
Li P~H~Y and Bishop R~F 2012 {\em Eur. Phys. J. B\/} {\bf 85} 25

\bibitem{Li:2012_anisotropic_kagomeSq}
Li P~H~Y, Bishop R~F, Campbell C~E, Farnell D~J~J, G{\"{o}}tze O and Richter J
  2012 {\em Phys.\ Rev.\ B\/} {\bf 86} 214403

\bibitem{Zheng:1991_honey}
Weihong Z, Oitmaa J and Hamer C~J 1991 {\em Phys. Rev. B\/} {\bf 44}
  11869--11881

\end{thebibliography}

\end{document}